\documentclass[preprint]{imsart}

\RequirePackage[OT1]{fontenc}
\RequirePackage{amsthm,amsmath}
\RequirePackage{natbib}
\RequirePackage[colorlinks,citecolor=blue,urlcolor=blue]{hyperref}
\arxiv{arXiv:1910.10102}
\usepackage{verbatim,color,amssymb,epsfig}
\usepackage{graphicx}
\usepackage{todonotes}
\startlocaldefs
\def\spacingset#1{\renewcommand{\baselinestretch}%
{#1}\small\normalsize} \spacingset{1}

\def\C{\bf{C}}
\def\X{\bf{X}}

\def\R{\bf{R}}

\def\t{\tau}
\def\balpha{\boldsymbol{\alpha}}
\def\bbeta{\boldsymbol{\beta}}

\def\bxi{\boldsymbol{\xi}}

\def\bse{\begin{eqnarray*}}
	\def\ese{\end{eqnarray*}}
\def\be{\begin{eqnarray}}
\def\ee{\end{eqnarray}}
\def\balpha{\boldsymbol{\alpha}}
\def\bbeta{\boldsymbol{\beta}}

\newtheorem{Th}{{\bf Theorem}}

\newtheorem{Rem}{{\bf Remark}}

\endlocaldefs

\begin{document}

\begin{frontmatter}

% "Title of the paper"
\title{Integrated Quantile RAnk Test (iQRAT) for gene-level associations}
\runtitle{Integrated Quantile RAnk Test }

% indicate corresponding author with \corref{}
% \author{\fnms{John} \snm{Smith}\corref{}\ead[label=e1]{smith@foo.com}\thanksref{t1}}
% \thankstext{t1}{Thanks to somebody} 
% \address{line 1\\ line 2\\ printead{e1}}
% \affiliation{Some University}
\begin{aug}
\author{\fnms{Tianying} \snm{Wang}\corref{}},
\author{\fnms{Iuliana} \snm{Ionita-Laza}}
\and
\author{\fnms{Ying} \snm{Wei}}
\runauthor{Wang, Ionita-Laza and Wei}
\affiliation{Center for Statistical Science, Department of Industrial Engineering, Tsinghua University \\
and\\
Department of Biostatistics, Columbia University}
\address{Center for Statistical Science, Department of Industrial Engineering, Tsinghua University,\\
Weiqing Building Rm 212-A, Beijing 100084, China\\
tianyingw@tsinghua.edu.cn}
\address{Department of Biostatistics, Columbia University, New York, NY 10032, USA\\
ii2135@cumc.columbia.edu\\
yw2148@cumc.columbia.edu}

\end{aug}
\begin{abstract}
Gene-based testing is a commonly employed strategy in many genetic association studies. Gene-trait associations can be complex due to underlying population heterogeneity, gene-environment interactions, and various other reasons. Existing gene-based tests, such as Burden and Sequence Kernel Association Tests (SKAT), are based on detecting differences in a single summary statistic, such as the mean or the variance, and may miss or underestimate higher-order associations that could be scientifically interesting. In this paper, we propose a new family of gene-level association tests which integrate quantile rank score processes to better accommodate complex associations. The resulting test statistics have multiple advantages: (1) they are almost as efficient as the best existing tests when the associations are homogeneous across quantile levels, and have improved efficiency for complex and heterogeneous associations, (2) they provide useful insights on risk stratification, (3)  the test statistics are distribution-free, and could hence accommodate a wide range of underlying distributions, and (4) they are computationally efficient. We established the asymptotic properties of the proposed tests under the null and alternative hypothesis and conducted large scale simulation studies to investigate their finite sample performance. We applied the proposed tests to the  Metabochip data to identify genetic associations with lipid traits and compared the results with those of the Burden and SKAT tests.
\end{abstract}

%\begin{keyword}[class=MSC]
%\kwd[Primary ]{}
%\kwd{Quantile process}
%\kwd[; secondary ]{}
%\end{keyword}


\end{frontmatter}

\section{Introduction} \label{sec:intro}

Gene-based association tests have important advantages over individual variant tests in GWAS analyses. By
directly identifying associated genes, they greatly improve functional interpretation. From the statistical perspective,
the number of tests is greatly reduced, which brings down the penalty for multiple testing, and leads to
more powerful tests. 
The increasing efficiency of generating large-scale genome sequencing datasets such as the data from the NHLBI Trans-Omics for Precision Medicine (TOPMed) Program \citep{taliun2019sequencing} also motivated the development of gene-based association tests \citep{morgenthaler2007strategy,li2008methods,morris2010evaluation,wu2011rare, wu2013kernel,chen2019efficient,he2019genome, ionita2011new}. As many variants identified in those studies have low population frequencies,  the primary test of interest is to test whether a group of variants within a region, such as a gene or noncoding region, are associated with a phenotype of interest. The existing  tests include the Burden tests and the sequence kernel association tests (SKAT).  The Burden tests aggregate information across the variants within a gene or region and then test for association between the resulting variant burden score and a phenotype of interest. Burden tests assume that genetic variants associated with the phenotype exhibit the same direction of association and have similar magnitude of effect \citep{bomba2017impact}. The SKAT tests \citep{wu2011rare, lee2012optimal} relax these assumptions by allowing a mixture of risk and protective variants, and allowing only a small percentage of causal variants in a region. Both tests are commonly used in the literature. {\color{black} Though many studies are focused on association tests with rare variants, tests for the joint effects of rare and common variants are desirable given the important contribution that common variants have to risk for complex traits and the current modest sample sizes for most sequencing studies \citep{han2010data,wang2015review}. Several such tests have been proposed in the literature, including the combined multivariate and collapsing (CMC) method \citep{li2008methods}, and extensions based on SKAT \citep{ionita2013sequence}. In this paper, we propose a gene-level quantile association test that helps identify the heterogeneous gene-trait associations. }

\paragraph{Heterogeneous Genetic Associations} Most existing tests evaluate whether genetic variants are associated with the mean of the phenotype \citep{madsen2009groupwise,morgenthaler2007strategy,wei2014detecting}, with only a few testing for effects on the variance \citep{schultz1985levene,brown2014genetic}.  However, genetic associations can be complex due to underlying heterogeneity in population and disease model, and the dynamic influence of gene-gene and gene-environment interactions \citep{manchia2013impact}.  Several recent studies have reported that genetic variants can influence other aspects of the phenotype distribution than the mean. For instance, \cite{yang2012fto} showed that a SNP in the FTO gene is not only associated with the mean of body mass index (BMI) but also with its variance. Similarly, variance quantitative trait loci (vQTLs) have been identified \citep{brown2014genetic,pare2010use,wang2019genotype}.
Identifying heterogeneous, higher order associations is a complementary way to make new genetic discoveries, and which can lead to more accurate risk stratification. 

Quantile-based approaches have been applied in several genetic studies  \citep{briollais2014application,beyerlein2011genetic}, which reported heterogeneous quantile-specific genetic effects on diverse complex traits.  For example,  \cite{beyerlein2011genetic} applied quantile regression to study the association between BMI and eight selected genetic variants, and found that their effect on childhood BMI is more pronounced among children with larger BMI. \cite{song2017qrank} also found that the eQTLs (expression quantitative trait loci) with heterogeneous quantile effects are associated with strong GWAS enrichment. Despite these significant findings for individual genetic variants, quantile-based associations have not been investigated for gene-based or set-based associations. 

\begin{figure}[!ht]
    \centering
    \includegraphics[scale = 0.2]{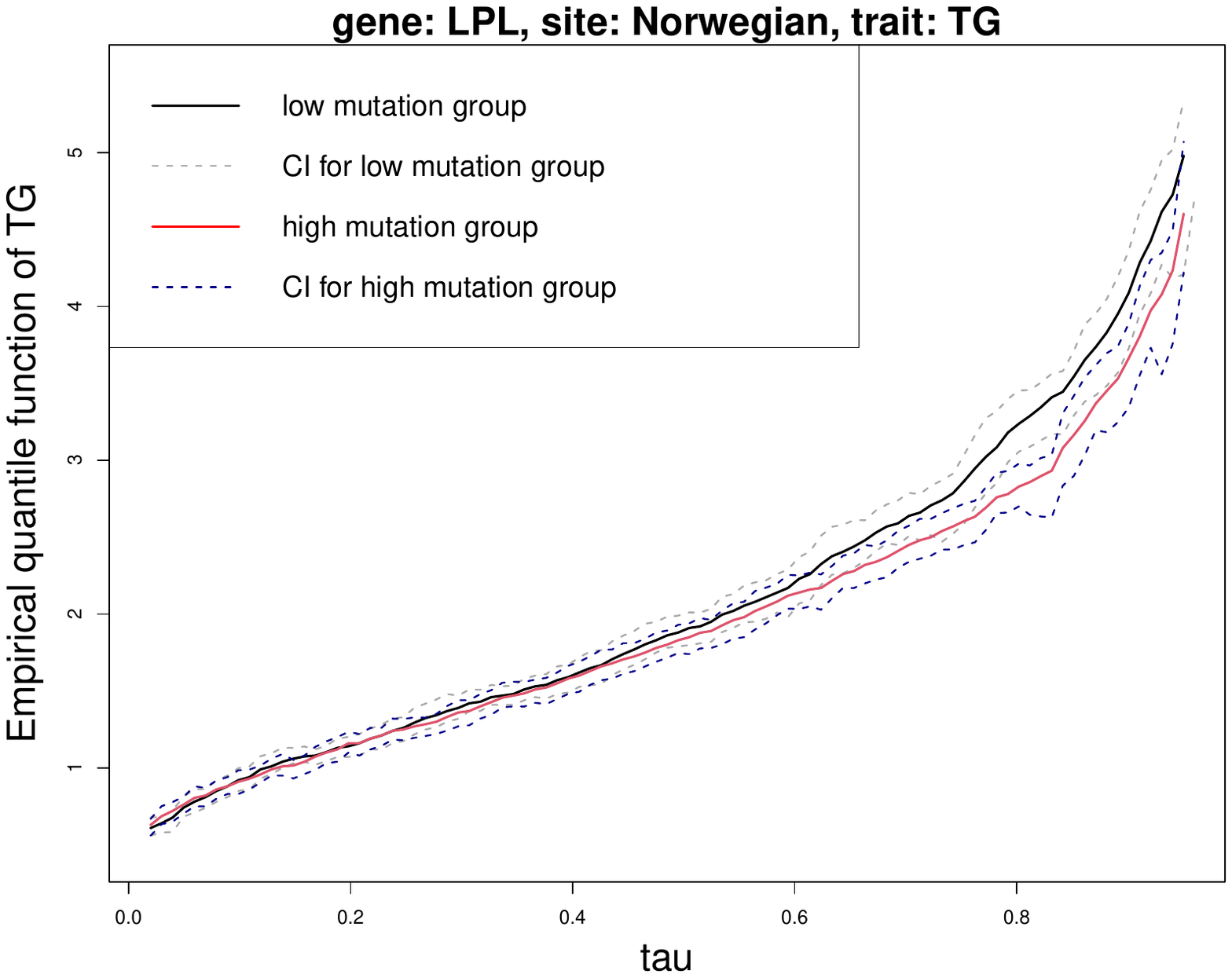}
 \includegraphics[scale = 0.2]{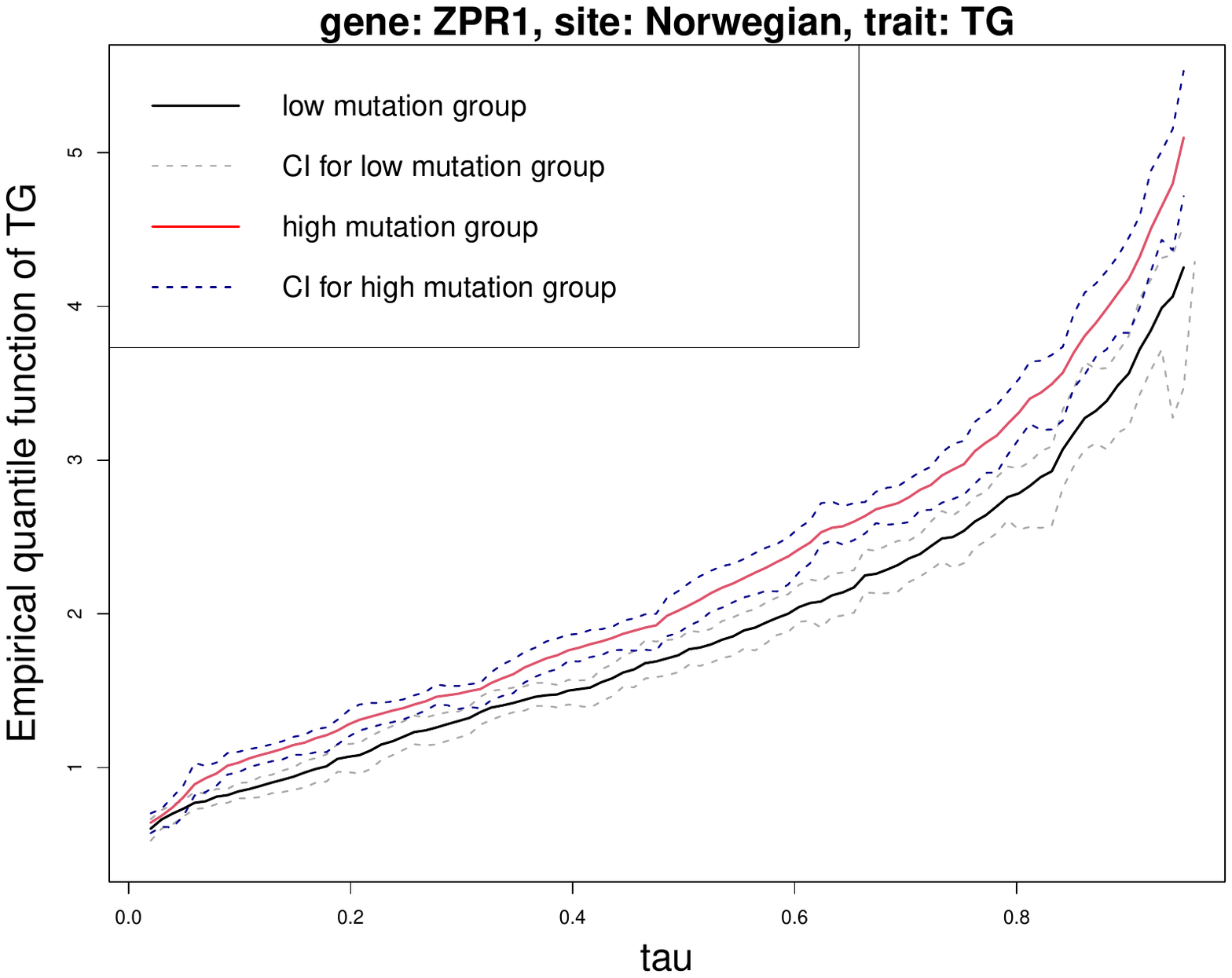}
 \caption{Empirical quantile function of $Y$ for genes LPL (\textbf{left}) and ZPR1 (\textbf{right}) in  Norwegian site. 95\% empirical confidence band is computed through bootstrap with 1000 replicates.} \label{fig10}
\end{figure}

{\color{black}
Such heterogeneous genetic associations were also found in the Metabochip study, which collected genotyping array data to assess associations with multiple traits \citep{voight2012metabochip}. We take the associations between genes and the lipid trait triglycerides as a motivating example.  %\todo[inline]{Add the brief description of Metabochip study.} 
To illustrate, we use the Norwegian sample in the Metabochip study with $n=2,793$. We select two genes {\it LPL} and {\it ZPR1}, which are among the top significant genes associated with triglycerides \citep{he2019genome,ference2019association,ueyama2015association,justice2018direct}. For each subject, we calculate the mutation burden for each of the two genes  LPL and ZPR1, i.e. $S_i=\sum_{j} w_jX_{ij}$, where $X_{ij}$ is  the $j$-th variant of $i$-th individual in the gene, and $w_j$ is the variant-specific weight  in \cite{ionita2013sequence}. We then stratify individuals according to the quartiles of $S_i$. We view the subjects in the lowest quartile ($\le 25\%$) as the low mutation group, and those in the top quartile ($\ge75\%$) as the high mutation group. We plot in Figure \ref{fig10} the empirical quantile curves of triglycerides among low and high mutation groups, overlaid with 95\% bootstrap confidence band. The difference between two empirical quantile functions describes how a gene impacts the different parts of the distribution.  We observed that {\it LPL} showed significant associations only at the upper quantiles, while {\it ZPR1} showed significant associations across all quantiles, with larger differences for larger quantile levels. }

Integrating such heterogeneity into testing could potentially increase the power, identify new genes, and provide useful insights on distributional differences. To this end, we proposed a new  Integrated Quantile RAnk Test(iQRAT)  to determine whether genetic variation within a gene leads to distributional differences in $Y$.

The proposed iQRAT uses quantile regression \citep{KoenkerBassett1978} to estimate the entire quantile process, and integrate its rank-score process\citep{gutenbrunner1993tests, koenker2010rank}  with various weighting schemes. Each weight scheme prioritizes a pattern of distributional difference that can be observed in genetic associations.   
These weighted test statistics are then combined for an overall gene-based association test. By construction, iQRAT is distribution-free. Hence, it generalizes the classical SKAT and Burden tests to  accommodate a wide range of distributions and more complex associations. They are also invariant to normalization transformations, which allows more direct interpretations.  We establish the asymptotic properties of iQRAT under both null and local alternative hypotheses, and  extensively compare both asymptotic efficiency and empirical power of the proposed iQRAT tests with existing approaches. In both theoretical and numerical investigations, we observed the enhanced power for detecting more complex and heterogeneous associations, especially when the target gene contains a mixture of common and rare variants.  When the data are normal with homogeneous association, the iQRAT tests are almost as efficient as the classical SKAT-based test.  In addition, as each weight function prioritizes certain distributional differences, post-hoc analyses on individual weighted test statistics can provide useful insights of the nature of gene-trait associations.

The rest of the paper is organized as follows: we present the proposed methodology and related asymptotic properties in Sections 2 and 3; in Section 4, we present a large scale simulation study to investigate the type I error and power under various models; in Section 5, we applied the proposed test in Metabochip data \citep{voight2012metabochip} to identify genes associated with lipid traits;  in Section 6, we discuss the advantages and limitations of the proposed method. Proofs for the asymptotic results and more plots are presented in the Supplementary Material.

\section{Methodology}

\subsection{Notations and background}
Throughout the paper, we denote a random sample as $(Y_i, \mathbf{X_i}, \mathbf{C_i}), i=1,...,n$, where $\mathbf{X_i}=(X_{i1}, \dots, X_{ip})$ is the $p$-dimensional genotype vector in a region (e.g., a gene)  for the $i$th individual,  $Y_i$ is the trait value, and $\mathbf{C_i}=(C_{i1}, \dots, C_{iq})$ is a $q$-dimensional covariate vector for the $i$th individual. The genotype vector $\mathbf{X_i}$ can be a mixture of both rare and common genetic variants. The goal is to determine whether  any of the $p$ genetic variants is associated with the outcome $Y_i$.  The classical linear model for genetic associations can be written as 
\begin{equation}\label{model}
E(Y_i|X_i, C_i)=\alpha_0+ \mathbf{C_i} \boldsymbol{\alpha} + \mathbf{X_i} \boldsymbol{\beta},
\end{equation}
where  $\boldsymbol{\beta}=(\beta_1, \dots, \beta_p)^\top$ are regression coefficients for the $p$ genetic variants. The hypothesis of interest is  $H_0:  \ \boldsymbol{\beta}=0,$ i.e. the mean of $Y_i$ is unrelated to $\mathbf{X_i}$.

To test $\boldsymbol{\beta}=0$ in eq \eqref{model}, the Burden and SKAT test statistics have been proposed \citep{wu2011rare, lee2012optimal,morgenthaler2007strategy,li2008methods}. They can be written in the form 
\begin{eqnarray*}%\label{q_ro}
Q_{\rho}&=&(\mathbf{Y}-\hat{\boldsymbol{\mu}}_0)^\top\mathbf{K_{\rho}}(\mathbf{Y}-\hat{\boldsymbol{\mu}}_0),
\end{eqnarray*}
where $\mathbf{Y}$ is the vector of the outcome $Y_i$,  $\hat{\boldsymbol{\mu}}_0$ is the vector of estimated means under the null model (i.e all $\beta$'s equal to zero),  $\mathbf{K_{\rho}}=\mathbf{X}\mathbf{WR_{\rho}W}\mathbf{X}^\top,$  $\mathbf{R_{\rho}}=(1-\rho)\mathbf{I}+\rho \mathbf{11^\top}$ specifies an exchangeable correlation matrix, and $\mathbf{W}=\textrm{diag}(w_1,\dots, w_p)$ is a diagonal weight matrix. 
The weights $w_1,\dots, w_p$ are pre-determined and assigned to each genetic variant. The choice of weights depends on individual application, according to the probability of these variants to be functional and hence more likely to influence the trait. By default, the weights are
inversely proportional to the minor allele frequencies (MAF) of the variants. Other functional scores such as CADD, DANN, FunSeq2, LINSIGHT, Eigen, FUN-LDA or DeepSEA can also be chosen \citep{kircher2014general,ionita2016spectral,quang2014dann,huang2017fast,backenroth2018fun,lu2016integrative,zhou2015predicting}. The Burden test  and the SKAT test  are special cases for $\rho=1$ and $\rho=0$, respectively. They can be written as
\begin{equation*}
Q_{\rm SKAT}=\sum_{j=1}^p w_j^2\left[\sum_{i=1}^n (Y_i-\hat{\mu}_{i,0})X_{ij}\right]^2, \\
Q_{\rm Burden}=\left[\sum_{j=1}^p w_j\sum_{i=1}^n (Y_i-\hat{\mu}_{i,0})X_{ij}\right]^2.
\end{equation*}
The null distribution of $Q_{\rm Burden}$ is a scaled $\chi_1^2$ distribution, and the null distribution of $Q_{\rm SKAT}$ follows a mixture of $\chi_1^2$ distributions. The $p$ values can be calculated based on the Davies method \citep{davies1980algorithm}.

\subsection{Proposed Integrated Quantile RAnk Test (iQRAT)}

To test the genetic association across quantiles, we extend the mean model \eqref{model} to the following  conditional quantile model of $Y$ given a genetic and covariate profile $(\mathbf{X}, \mathbf{C})$,  
\begin{equation} \label{eq:main}
Q_{Y_i}(\t | {\C_i}, {\X_i}) = \alpha_0(\tau) +\C_i^\top\boldsymbol{\alpha}(\tau)+ \X_i^\top\boldsymbol{\beta}( \tau), \forall \tau\in(0,1),
\end{equation}
where $\bbeta(\tau)=(\beta_1(\tau), \beta_2(\tau),...,\beta_p(\tau) )^\top$ is the $p$-dimensional quantile coefficient functions associated with the gene $X_i$,  $\boldsymbol{\alpha}(\tau)$ are those associated with the covariate $\C_i$, $\alpha_0(\tau)$ is the intercept function.  One can view  $\alpha_0(\tau)$ as the quantile function of $Y$ when both $\mathbf{X}$ and $\mathbf{C}$ are zero.  In the rest of the paper, we call $F(\cdot) = \alpha_0^{-1}(\tau)$ the error distribution of Model \eqref{eq:main}.

Next, we propose a new group-wise quantile association test to test the hypothesis $$H_0:\bbeta(\tau)=  0,\,\, \forall\ \tau \in (0,1),$$   i.e., whether the quantile function of $Y$ is related to the genotypes $\mathbf{X}$ at any quantile level $\tau\in(0,1)$.  We call the proposed test  Integrated Quantile RAnk Test (iQRAT). We construct iQRAT as follows:

\begin{description}

\item[Step 1: ] \textbf{Estimate the conditional quantile process under the null model, and construct individual  quantile rank score processes accordingly.} Under the null hypothesis $\bbeta(\tau)=  0$, the conditional quantile of $Y$ given $\mathbf{X}$ and $\C$ can be written as  $Q_Y(\tau | \mathbf{X}, \C) = \C^\top\balpha(\tau)$. We use quantile regression to regress $Y$ against $\C$ over the entire quantile process, and denote the resulting estimates as  $\widehat{\balpha}_{\rm null} (\tau)$. We refer to \cite{koenker1990note,gutenbrunner1993tests,koenker2014convex, wei2009quantile} for technical details of  quantile process estimation.

%by
%\begin{equation*}
%    \widehat{\balpha} (\tau) = \arg\min_{\balpha} %\sum_{i=1}^n {\large{\boldsymbol{\rho}}_{\tau}}(Y_i %- {\C_i}^\top\balpha), 
%\end{equation*}
%where ${{\boldsymbol{\rho}}_{\tau}} (u) =  \tau|u|I\{u>0\} + (1-\tau)|u|I\{u\le0\}$ is the quantile regression loss function, and $(\tau_1,...,\tau_{k_n})$ are $k_n$ estimable quantile levels. We use parametric linear programming to estimate the entire quantile process with all the $k_n$ estimable quantile levels given the data.  Similar as in , we then estimate the quantile coefficient function $\balpha(\tau)$ by a linear spline expanded from the $\{\widehat{\balpha}_{\tau_k}, k=1,...,k_n\}$, and denote it as $\widehat{\balpha}_n(\tau)$. Next, we construct a quantile regression rank score process by
For each observation, we define its rank-score process (under the null) by 
$$\widehat{a}_i(\tau) = \mathbf{1}\{Y_i < \C_i\widehat{\balpha}_{\rm null}(\tau)\}-\tau,$$  where $\mathbf{1}\{Y_i < \C_i\widehat{\balpha}_{\rm null}(\tau)\}$ is a binary indicator whether $Y_i$ stays underneath the $\tau$-th estimated conditional quantile. If the null hypothesis $\bbeta(\tau)=0$ is true, we  expect the score function  $E(\widehat{a}_i(\tau))=0$ for any $\tau \in (0,1)$. A deviation from zero at any quantile level $\tau$ suggests the existence of genetic associations.

\item[Step 2: Integrate $\widehat{a}_i(\tau)$ over $\tau$ with multiple weight functions.] As $\widehat{a}_i(\tau)$ indicates quantile-specific associations, a natural way to measure the overall genetic association is to integrate $\widehat{a}_i(\tau)$ over quantile levels $\tau$. We consider weighted integrations to enhance the detection of heterogeneous associations.

Let $\varphi: (0,1)\to \R$ be a non-decreasing square-integrable function.  We integrate each
$\widehat{a}_i(\tau)$ over $\tau$ with respect to the $\varphi(\cdot)$ 
by
$$
    \widehat{\phi}_i^\varphi =\int_0^1\widehat{a}_i(\tau)d\varphi(\tau) ,\ \  i = 1,...,n.
$$
%Since $\widehat{a}_i(\tau)$ is a piece-wise linear function, $\widehat{\phi}_i^\varphi$ can be written as
%\begin{equation*}
%    \widehat{\phi}_i^\varphi = \int_0^1 \widehat{a}_i(t)d \varphi(t) = \sum_{j=1}^{k_n} \frac{\widehat{a}_i(\tau_j)-\widehat{a}_i(\tau_{j-1})}{\tau_j - \tau_{j-1}} \int_{\tau_{j-1}}^{\tau_j} \varphi (t)dt.
%\end{equation*}

The integrated rank score $\widehat{\phi}^\varphi$ essentially accumulates the evidence across quantile levels, and uses  the first derivative $\partial\varphi(\tau)/\partial \tau$ as the weight function to assign different weights at different quantile levels.

When $\varphi(\tau)$ is a linear function of $t$, the resulting  rank score $\widehat{\phi}^\varphi$ is an unweighted average over quantile process, and hence is equivalent to the mean effect. Following the notations in \citep{gutenbrunner1992regression,gutenbrunner1993tests, koenker2010rank}, we call $\varphi_1(\tau)=\tau$ as the Wilcoxon weighting function. Several studies, including \cite{zou2008composite, kai2010local}, estimated the mean by averaging quantile functions, and found that it leads to more efficient  mean estimation than the classical least square estimators in the presence of non-normal errors.

Besides $\varphi_1(\tau)=\tau$, we also consider the following three weight functions: (1) The Normal weighting function: $\varphi_2(\tau) = \Phi^{-1}(\tau)$, where $\Phi(\cdot)$ is the standard normal distribution function,  (2) The Lehmann weighting function: $\varphi_3(\tau) = -\log(1-\tau)-1$, and {\color{black}(3) Inverse-Lehmann weighting function: $\varphi_4(\tau) = \log(\tau)+1$.}

\begin{figure}[!ht]
\begin{center}
    \includegraphics[height = 1.4in, width = 2.25in]{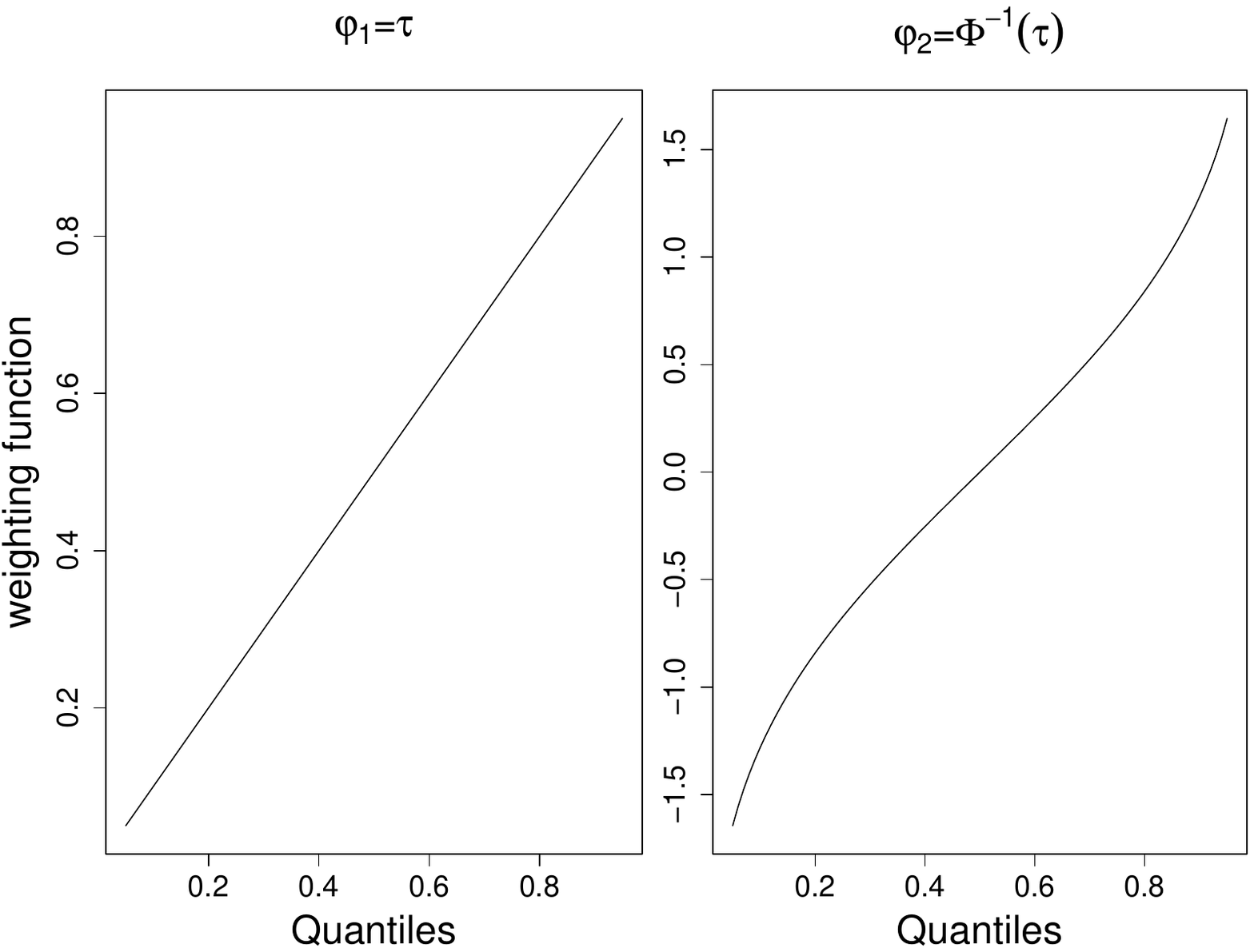}
     \includegraphics[height = 1.4in, width = 2.25in]{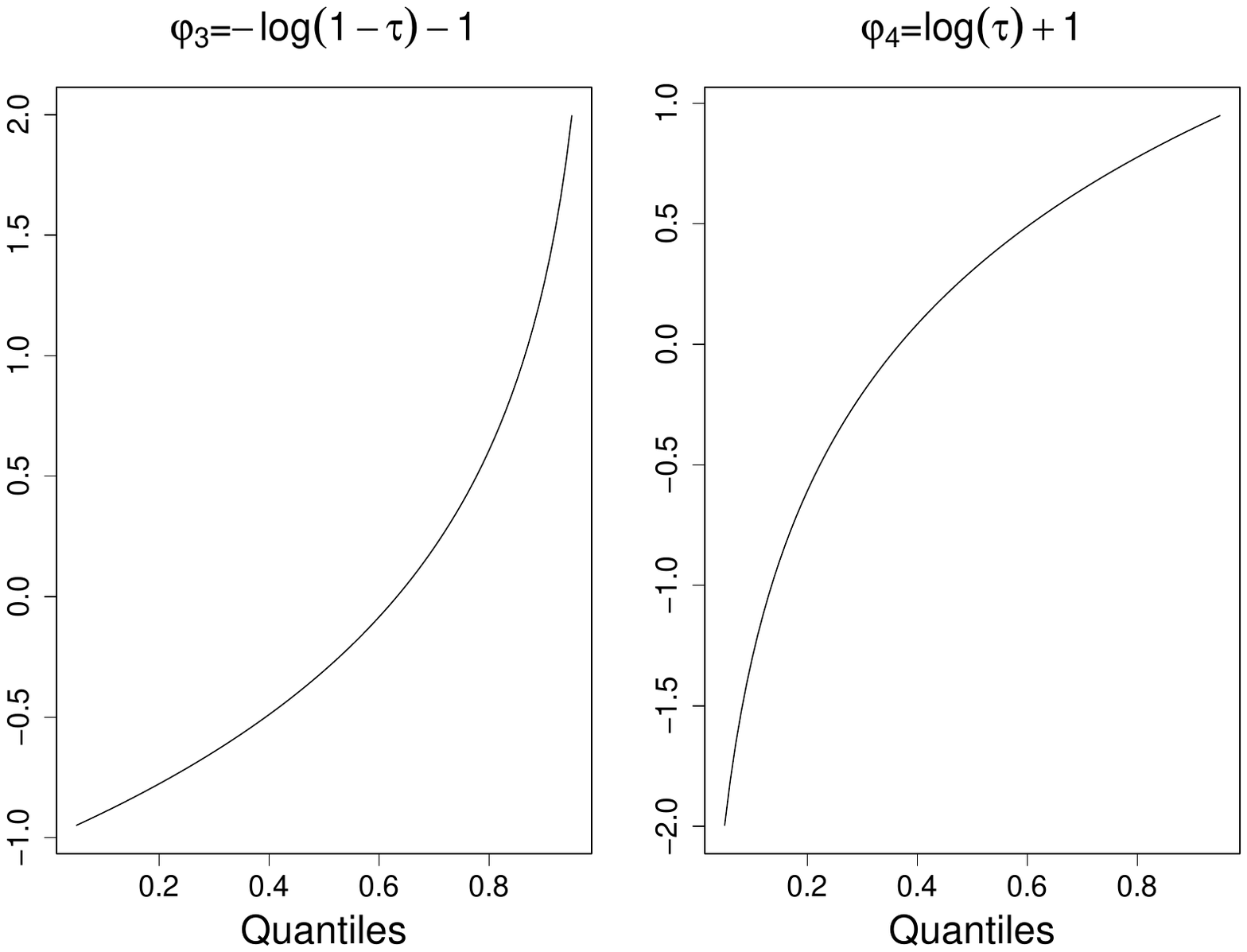}
     \includegraphics[height = 1.4in, width = 2.25in]{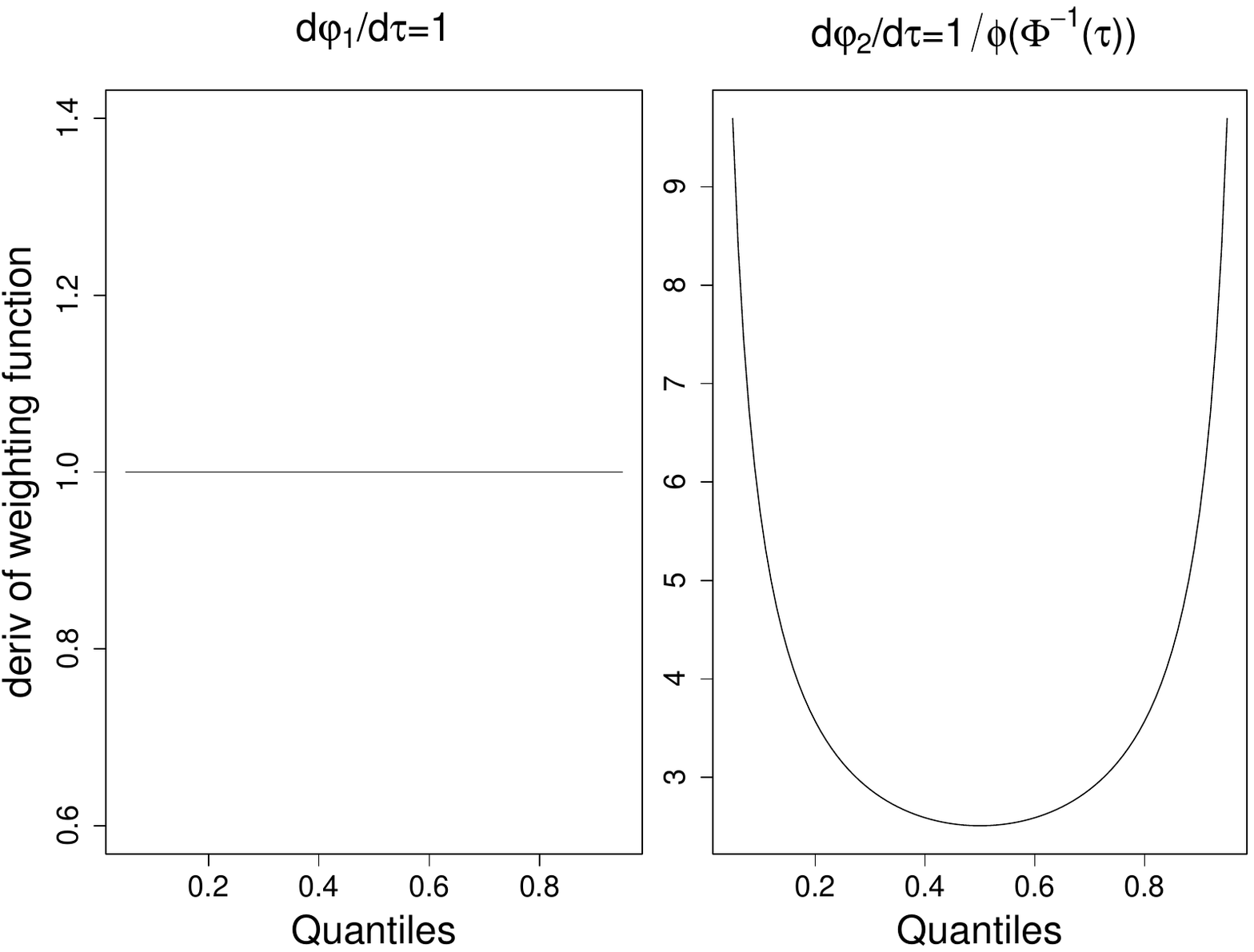}
     \includegraphics[height = 1.4in, width = 2.25in]{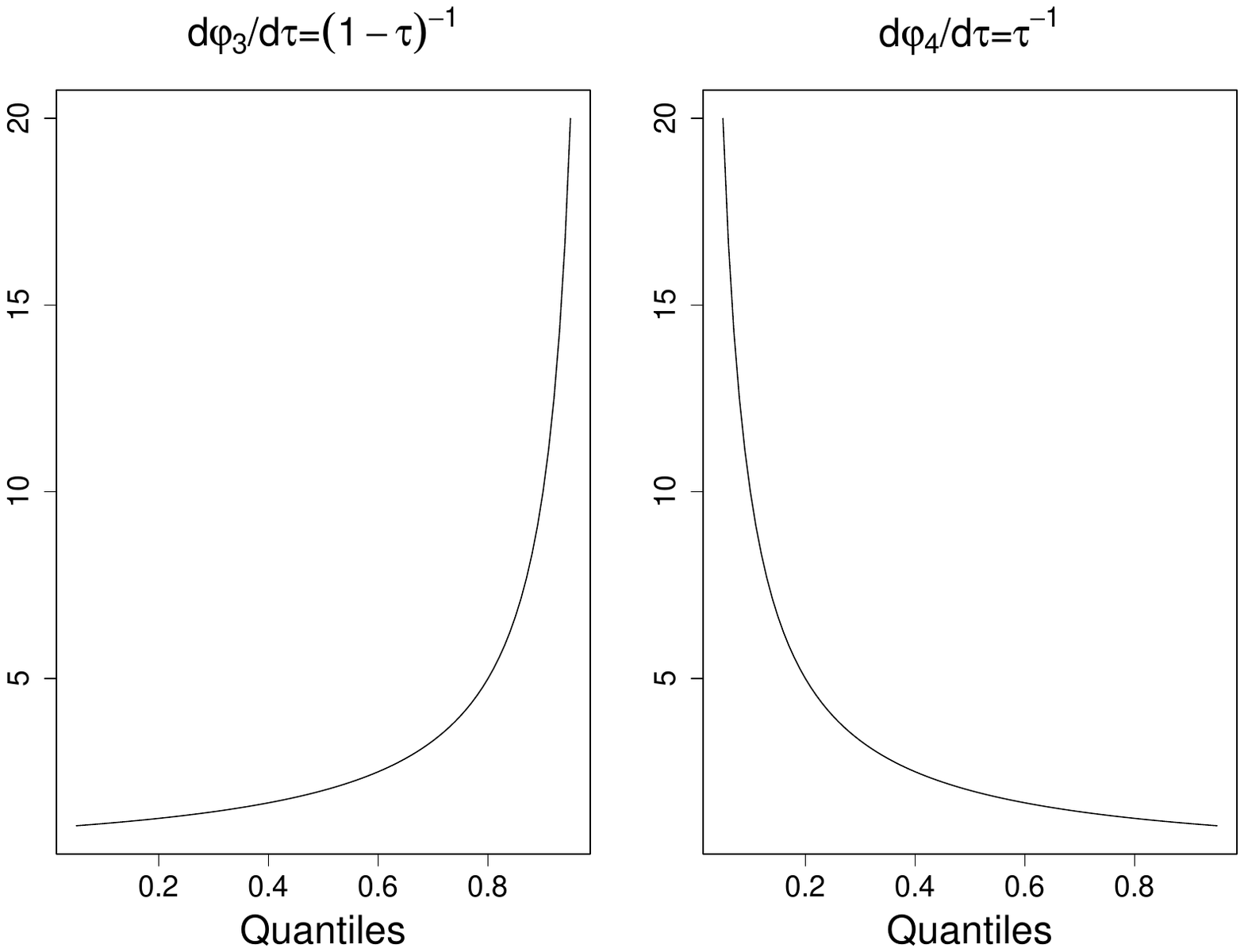}
\caption{The quantile weighting functions $\varphi$ (\textbf{top}) and their first derivatives $d\varphi/d\tau$ (\textbf{bottom}): $\varphi_1$(Wilcoxon), $\varphi_2$(Normal),  $\varphi_3$(Lehmann) and $\varphi_4$(Inverse Lehmann).}\label{fig:phi4}
\end{center}
\end{figure}

Figure \ref{fig:phi4} displays the four weight functions and their first derivatives, $\varphi_1(\tau)$ (Wilcoxon weighting), $\varphi_2(\tau)$ (Normal weighting), $\varphi_3(\tau)$ (Lehmann weighting) and $\varphi_4(\tau)$ (Inverse-Lehmann weighting). 
As shown, $\partial\varphi_2(\tau)/\partial \tau$ is symmetric around the median with heavier weights at the two tails. The resulting integrated rank score is  asymptotically optimal for Gaussian error under the location shift model \citep{gutenbrunner1993tests}.
On the other hand, the first derivatives of Lehmann and Inverse Lehmann weight functions are asymmetric. The Lehmann weights $\partial\varphi_3(\tau)/\partial \tau$ assign increasingly higher weights at the upper tail. As a result, it is optimal to detect distributional differences at upper tails. In contrast, the Inverse Lehmann weight function focuses on the differences at the lower tails.

%\todo{In what setting?}{\color{red}Under the location shift alternative hypothesis},  Normal score is asymptotically optimal when the error $\epsilon$ follows a normal distribution; Wilcoxon score is optimal for logistic distributed error \citep{gutenbrunner1993tests, koenker2010rank}. Lehmann score is optimal for Lehmann alternative, a combination of global location and scale shift with a strong local effect on the right tail.

\item[Step 3: Construct iQRAT test statistics for each $\varphi$.]  For each weighting function $\varphi$, we construct the following test statistics:
\begin{equation*} \label{eq:keyteststat}
{\bf S}^\varphi = n^{-1/2}\sum_{i=1}^n \X^{*\top}_{i}\widehat\phi_i^\varphi,
\end{equation*}
where $\X^*_i$ is the genotype vector after being orthogonalized against the covariate matrix. Let $\C_n$ be the $n\times q$ design matrix associated with the covariates, and $\mathbf{P}_{\C} = \C_n(\C_n^\top \C_n)^{-1}\C_n^\top$ is the projection matrix onto the linear space of $\C_n$. $\X^*_i$ is the $i^{th}$ row of  the matrix $\X^{*} =(I - P_C )\X_n $, where $\X_n$ is the $n\times p$ design matrix associated with the genotypes. The orthogonalization ensures the asymptotic independence between the genetic association and covariates.  The test statistic  $\mathbf{S^\varphi}$ is in the category of rank-based statistics \citep{sidak1999theory,gutenbrunner1993tests}. We propose two integrated quantile rank test (iQRAT) statistics that generalize the SKAT and Burden tests in the following forms: 
 \begin{eqnarray}
Q_S^{\varphi} &=& \mathbf{S}^{\varphi \top} W^2\mathbf{S}^\varphi =  \sum_{j=1}^pw_j^2\left(\sum_{i=1}^n \phi_i^\varphi X_{ij}^*\right)^2,\label{QS} \\ 
Q_B^\varphi &=& \mathbf{S}^{\varphi\top} W1_p 1_p^\top W \mathbf{S}^\varphi =  \left(\sum_{j=1}^pw_j\sum_{i=1}^n \phi_i^\varphi X_{ij}^*\right)^2,\label{QB} 
 \end{eqnarray}
 where $W = {\rm diag}(w_1,...,w_p)$ is the diagonal weight matrix for $p$ individual genetic variants. The weights $w_j$'s  are pre-determined, and measure the relative likelihood of the $j$th genetic variant to be functional.  We discuss the details of the choice of $W$ in the subsequent section \ref{sec:combine}. 
 In section \ref{sec:asymptotics}, we establish the asymptotic distributions of $Q_S^{\varphi}$ and $Q_B^\varphi$ under both the null and alternative hypotheses.

\item[Step 4: Combine $\varphi$-specific tests into the final iQRAT test.] 
 As each quantile weighting function captures a certain type of association pattern, we  propose to integrate the rank-score process using each of the four $\varphi(\tau)$ functions, and then 
 use  the Cauchy combination test recently proposed in \cite{liu2018cauchy}   
 combine their $p$-values into the final iQRAT test. Let $p_1,...,p_k$ be $k$ $p$-values, which follow a uniform (0,1) distribution under the null hypothesis. The Cauchy p value combination method combines them by computing $\sum_{i=1}^k\tan\{(0.5-p_i)\pi\}/k$. One can show that
 $\tan\{(0.5-p_i)\pi\}$ follows a standard Cauchy distribution for any $i$. Consequently, $\sum_{i=1}^k\tan\{(0.5-p_i)\pi\}/k$  is also a standard Cauchy distribution for any $k$. In other words,  the test correlations have limited effect on the tail distribution of Cauchy combined $p$-values, and  we easily use the standard Cauchy distribution to determine the overall $p$-value of the combined statistic. 
 The Cauchy combination method is computationally simple and allows the combined tests to be correlated.  The unified test statistic shows robust power improvement, while the test statistic with single quantile weighting function can provide useful insights into the possible local association patterns.
\end{description}

{\color{black}
 \begin{Rem}
 The test statistics $Q_S^{\varphi}$ and $Q_B^\varphi$ are in the category of rank-score test, but are distinct from the existing rank score tests in quantile regression \citep{koenker2010rank}. Due to the existence of rare variants, the covariance matrix of $S$ is nearly singular. Hence the classical rank-score test in \citep{koenker2010rank, gutenbrunner1993tests} and its multivariate version in \cite{song2017qrank} cannot be applied directly. The asymptotic and empirical properties need to be investigated separately. 
 \end{Rem}
 \begin{Rem} There are several existing approaches in literature to combine multiple $p$-values, such as the Fisher's method, minimum $p$-value, higher criticism, Berk-Jones \citep{fisher1992statistical, dudoit2003multiple, jin2006higher, moscovich2016exact,sun2019powerful}. In our approach, the $p$-values from the same set of variants with different score functions $\varphi(\tau)$ are highly correlated. These traditional approaches for combining $p$-values require resampling or permutation to estimate the correlations, which are computationally expensive. That is why we employ here the Cauchy combination method.
 \end{Rem}
}

 \subsection{iQRAT test with Variants Stratification} \label{sec:combine}
 
 In this section, we discuss the practical consideration of implementing the proposed iQRAT, which includes (1) the stratification of common and rare variants, and (2) the rationale behind the choice of four quantile weighting functions.

 \paragraph{Variants Stratification}

In gene-based association tests, assigning weights to individual variants is a common strategy to enhance the power of the test by leveraging  external knowledge \citep{wu2011rare, ionita2013sequence,madsen2009groupwise}. The weights are often chosen to be inversely proportional to the MAFs. The underlying rationale is that rare or low frequency variants are more likely to be disease associated.  

For many complex traits, risk variants may range from rare to common \citep{li2008methods}. Several studies  \citep{wu2011rare,lee2012optimal,jeng2016rare, bomba2017impact}
reported that a single MAF-based weighting scheme often over-penalizes the common variants, and in turn undermines the detection of gene-level associations. For this reason, we follow the recommendations in \cite{ionita2013sequence} to stratify the variants into rare and common groups. 

\begin{figure}[!ht]
    \centering
    \includegraphics[scale = 0.55]{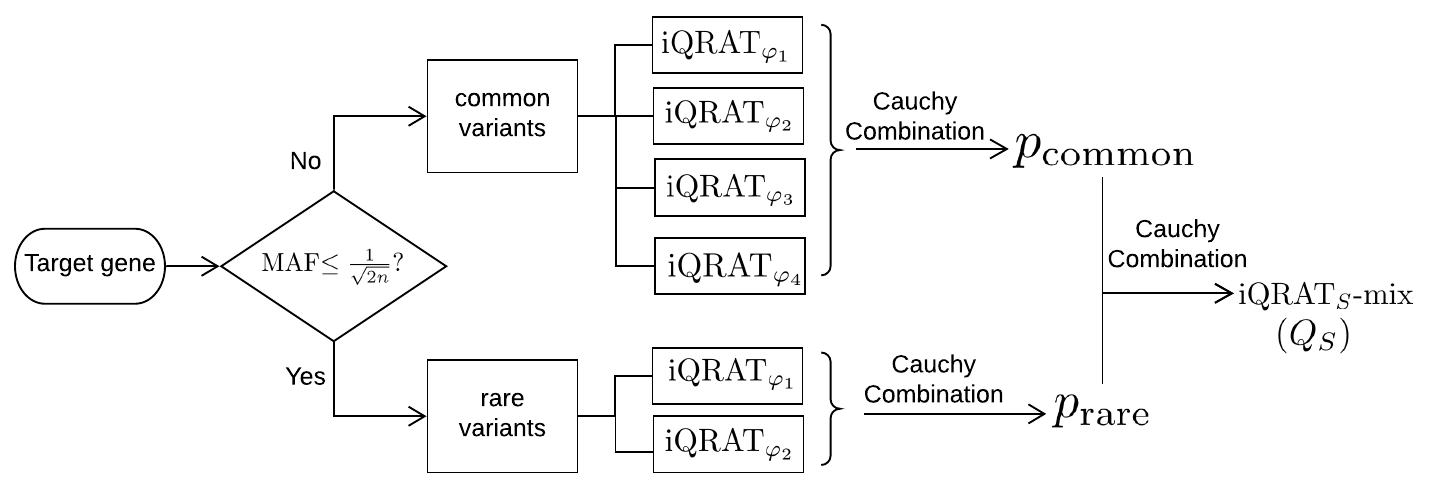}
    \caption{Implementation procedure for iQRAT test for rare and common variants.}
    \label{fig:flowchart}
\end{figure}

Let $p_j$ be the sample MAF  of the $j$th variant in a target gene. We use  an adaptive threshold $1/\sqrt{2n}$, where $n$ is the total sample size, to stratify the variants. Specifically, 
we assign a variant to the common group if  $p_j>1/\sqrt{2n}$, and assign a variant to the rare group, if 
 $p_j \le 1/\sqrt{2n}$. After the partition, we construct the iQRAT test statistics separately for the common and rare variants. For common variants, we construct the iQRAT test statistics using all the four score functions $\varphi(\tau)$ respectively, and using the variant weights $w_j= Beta(p_j,0.5,0.5)$  
where $Beta$ stands for the density function of a Beta-distribution. Following the outlined procedure in the precedent section, we use Cauchy Combination to combine the p-values from the four $\varphi$ specific iQRAT statistics. We denote the resulting p-value as $p_{\rm common}$. 
For rare variants, we only construct the iQRAT  statistics using the Normal ($\varphi_2$) and Wilcoxon ($\varphi_1$) score functions, and use the variant weights
$w_j=  Beta(p_j,1,25)$. We also use the Cauchy combination method to obtain the p-value for rare variants, $p_{\rm rare}$. {\color{black}Finally, we use the Cauchy combination to combine $p_{\rm common}$ and $p_{\rm rare}$ into the final p-value for the target gene. Figure \ref{fig:flowchart} displays the flow chart of the proposed iQRAT test procedure. }
%\todo[inline]{add flow chart here}
{\color{black}
\paragraph{Rationale behind the choice of weights for the rare variant test} The Lehmann and Inverse Lehmann score functions are designed to prioritize the tail differences at extreme quantiles. In the rare variant group, we often do not have enough carriers of such rare variants.  As a result, we do not have sufficient samples to detect a tail difference. Incorporating those score functions into the rare variant tests could lead to variance-inflation, and increased false positive rates. Same as in \cite{ionita2013sequence}, we  used different variant weights for common and rare variants, since they are optimized  for common and rare variants, respectively. 
Instead of combining all the 6 $p$-values (4 for common variants, and 2 for rare variants) at once, we used two-stage combinations. We first combine the p-values within common and rare groups separately, and then combine  $p_{\rm common}$ and $p_{\rm rare}$ in the second stage. Such design is to ensure the equal contribution of rare and common variants.}

\section{Asymptotic results}\label{sec:asymptotics}

\subsection{Asymptotic distributions for $Q_B$ and $Q_S$}
 In this section, we establish the asymptotic distributions for the test statistics $Q_S^{\varphi}$ and $Q_B^{\varphi}$ respectively under  the null hypothesis  and a set of local alternatives. 

The two iQRAT test statistics $Q_S^{\varphi}$ and $Q_B^{\varphi}$,  as defined in eq\eqref{QS}-\eqref{QB}, are built upon the rank-score statistics $
{\bf S}^\varphi = n^{-1/2}\sum_{i=1}^n \X^{*\top}_{i}\widehat\phi^\varphi
$.  We first establish the asymptotic normality of 
${\bf S}^\varphi$ in the following theorem. To do so, we make
 a few assumptions. We assume that the errors are independent and identically distributed with an absolutely continuous density $f$. The quantile weighting function $\varphi$ is nondecreasing and square-integratable over $(0,1)$.  We also impose some mild conditions on the design matrix $(\mathbf{1,C})$ to obtain a valid Bahadur representation of regression quantiles.  We outline the detailed conditions in Supplementary Material, see Conditions A-C. They are consistent with the quantile rank score literature \citep{gutenbrunner1993tests}. 
Under those conditions, we establish the asymptotic normality of 
${\bf S}^\varphi$ in the following theorem, and derive  the asymptotic distributions of $Q_S^{\varphi}$ and $Q_B^{\varphi}$ accordingly. 
For simplicity, we define $\Sigma = n^{-1} \X^{*\top}\X^{*}$, i.e. the component that does not depend on $\varphi(\cdot)$ and the error distribution.

%With the outlined conditions in the Supplementary Material, the  statistics  ${\bf S}^\varphi$ is asymptotically normally distributed. Following the asymptotic normality of $\mathbf{S}^\varphi$, we derive the asymptotic distributions of $Q_S^{\varphi}$ and $Q_B^{\varphi}$. 

 {\color{black} }

 %distribution of  under $H_0$ is a normal distribution with mean zero; while under the alternative distribution, it has mean $\gamma(\varphi, F)\Sigma^\varphi\beta_0$, where $\gamma(\varphi, F) = -\int_0^1\varphi(t)df(F^{-1}(t))$, $\beta_0$ is the fixed vector in $H_n$. Then, in Theorem \ref{Thm:1} and \ref{Thm:2}, we present the asymptotic distribution of $Q_S$ and $Q_B$ under the null hypothesis and the alternative hypothesis, separately. The theoretical results are based on simple local alternatives $H_n: \beta_\tau = n^{-1/2}\beta_0, \ \tau\in(0,1)$. Proofs for all theorems and supportive Lemmas are stated in the Appendix A. For 

\paragraph{Under the null hypothesis}

\begin{Th}\label{Thm:1}

Under the conditions A-C (in the Supplementary Material), and under the null hypothesis
$H_0: \boldsymbol{\beta}(\tau) = 0$, we have 
\begin{enumerate}
\item $\mathbf{S}^\varphi$ follows asymptotically a normal distribution
$\mathbf{S}^\varphi = AN(0, \sigma^2_\varphi\Sigma ),$
where  $\sigma^2_\varphi = \int (\varphi(t) - \bar{\varphi})^2dt$ and $\bar{\varphi} = \int_0^1\varphi(t)dt.$
    \item $Q_S^{\varphi} $ is asymptotically a mixture of $\chi^2_1$ distributions: $Q_S^{\varphi} = \sigma^2_\varphi\sum_j^p \lambda_j \chi^2_1,$ 
where $\lambda_j for j = 1,...,m$ are positive eigenvalues of $\Sigma^{1/2}W^2\Sigma^{1/2}$; If  $\Sigma^{1/2}W^2\Sigma^{1/2}$ is semi-positive definite, we sum over the first $p$ positive eigenvalues instead of all $m$ eigenvalues.
\item $Q_B^{\varphi} $ follows a scaled $\chi^2_1$ distribution: $Q_B= \sigma^2_\varphi \lambda \chi^2_1,$ 
 where $\lambda = 1^\top_p W\Sigma W 1_p$.
\end{enumerate}

\end{Th}
The rank-score statistics  $\mathbf{S}^\varphi$ is \textit{distribution-free} in the sense that its asymptotic distribution under the null hypothesis only depends on the score/weight function $\varphi(t)$ and the design matrix. This feature makes it flexible to accommodate a wide range of trait distributions. {\color{black} The $p$ value of $Q_S^\varphi$ can be approximated efficiently using Davies method based on the numerical inversion of the characteristic function \citep{davies1980algorithm}.
}

\paragraph{Under the alternative hypothesis}
When $\bbeta(\tau) \neq 0$, the test statistics $Q_S^\varphi$ and $Q_B^\varphi$ have no longer mean zero. Their non-central parameters $\eta$ depend on the form of alternatives $\bbeta(\tau)$, error distribution $F$ and the weight score function $\varphi(\tau)$.  Theorem \ref{Thm:2}  presents the asymptotic distributions of $Q_S^\varphi$ and $Q_B^\varphi$ under alternatives. %, which depends on $\bbeta(\tau)$ and the error distribution $F() = \alpha_0^{-1}(\tau)$

\begin{Th}\label{Thm:2}  Under the conditions A-C (in the Supplementary Material),  we have 

\begin{enumerate}
    \item $\mathbf{S}^\varphi$ follows asymptotically a normal distribution  $
\mathbf{S}^\varphi = AN(\bxi^\top\Sigma,\  \sigma^2_\varphi\Sigma)
$, where $\bxi = (\xi_1,...,\xi_p)$, and $\xi_j= \int_0^1f(F^{-1}(\tau))\beta_j(\tau)d\varphi(\tau)$ for $j = 1,...,p$;

    \item The distribution of $Q_S^{\varphi}$ converges to a linear combination of non-central chi-square distributions 
    $Q_S^{\varphi} \xrightarrow[]{d} \sum_j^m \sigma^2_\varphi\lambda_j \chi^2_1(\eta_j),$
    where $\lambda_j$'s are the positive eigenvalues of $\Sigma^{1/2}W^2\Sigma^{1/2}$ and $\eta_j$'s are non-central parameters. Let $U$ be an orthonormal matrix which satisfies $\Lambda = U\Sigma^{1/2}W^2\Sigma^{1/2}U^\top$ and
    $\Lambda_{p\times p}=diag(\lambda_1,...,\lambda_m,0,...,0)$. We can write the non-central parameters  $\eta_j= \mu_j^2$  where $\mu_j$ is the $j^{th}$ element of $\mu = U\Sigma^{-1/2} \bxi/\sigma_\varphi$.

     \item   The distribution of $Q_B^{\varphi}$ converges to a scaled non-central chi-square distribution
     $Q_B= \lambda \chi^2_1(\eta ),$
   where $\eta = \bxi^\top \Sigma W 1_p 1^\top_pW \Sigma \bxi$ and $\lambda = \sigma^2_\varphi 1^\top_pW\Sigma W 1_p$.
    \end{enumerate}

\end{Th}
Proofs for Theorems \ref{Thm:1}-\ref{Thm:2} can be found in the Supplementary Material.

In theory, one can choose an optimal $\varphi(\tau)$ by maximizing the non-central parameter. However, the non-central parameters depend on actual $\bbeta(\tau)$ and $F$, which are often unknown and could be very different across genes.  Hence, it is hard to identify a simple $\varphi(\tau)$ that works for all genes. Adaptive $\varphi(\tau)$ is appealing but often numerically challenging. Hence combining multiple pre-determined but representative weight functions $\varphi$ is a more practical strategy  to accommodate complex associations and to enhance the statistical power.

\subsection{Asymptotic efficiency of the iQRAT tests}   
In this section, we compare the  asymptotic efficiency of the proposed iQRAT tests with their mean-based counterparts under various alternative settings. As we derived in Theorem \ref{Thm:2},  the asymptotic distribution of $Q_\varphi$ is the same as the distribution of  $\sigma^2_\varphi\sum_j\lambda_j\chi^2_1(\eta_j)$,  where $\chi^2_1(\eta_j)$'s are independent non-central chi-square distributions with non-central parameters $\eta_j$ and degree-of-freedom 1. The non-central parameters
$\eta_j =  \bxi^\top u_j^\top u_j \bxi/(\sigma^2_\varphi \lambda_j)$, where $\lambda_j$ and $u_j$ only depend on $\X$,   $\sigma^2_\varphi = \int (\varphi(t) - \bar{\varphi})^2dt$, $\bar{\varphi} = \int_0^1\varphi(t)dt$, $\bxi = (\xi_1,...,\xi_p)$ and $\xi_j= \int_0^1f(F^{-1}(\tau))\beta_j(\tau)d\varphi(\tau)$ for $j = 1,...,p$. On the other hand, 
the asymptotic distribution of $Q_{\rm SKAT}$ shares the same form, except that $\bxi$ is replaced by $\overline{\bbeta} = (\overline{\beta_1},...,\overline{\beta_p})$ and $\overline{\beta}_j = \int_0^1 \beta_j(\tau) d\tau$ for $j = 1,...,p$, and 
$\sigma^2_\varphi$ is replaced by $\sigma^2 = \overline\beta^2 \sigma^2_x+\sigma^2_e=\overline\beta^2 + \sigma^2_e$. 

 To this end, we define $\text{eff}(T) = \mu^2(T)/{\rm Var}(T)$ as the efficiency measure of a test statistics $T$, where  $\mu(T)$ and ${\rm Var}(T)$ are its asymptotic mean and variance. Without loss of generality, we assume  that $\mathbf{X}$ is univariate with  variance 1. Similar results hold for multiple dimensional $\mathbf{X}$. In this univariate setting, SKAT-type test statistics are equivalent to Burden-type tests, and we denote them both as $Q_{\rm mean}$. It follows that we can write the efficiency $\text{eff}(Q_\varphi)$ and $\text{eff}(Q_{\rm mean})$ as
\bse 
\text{eff}(Q_\varphi) = \frac{(1+\xi^2/\sigma^{2}_\varphi)^2}{2(1+2\xi^2/\sigma^{2}_\varphi)} \ \ \mbox{and} \ \ \text{eff}(Q_{\rm mean}) = \frac{(1+\bar\beta^2/\sigma^{2})^2}{2(1+2\bar\beta^2/\sigma^{2})}.
\ese 
The efficiency depends on the alternative hypothesis and its corresponding quantile effect $\beta(\tau)$, the error distribution $F(\cdot)$ and the quantile weighting  function $\varphi(\tau)$. Since normalization is a common practice in genetic association tests, we consider $F(\cdot)$ as a standard normal distribution in this section. Empirical power comparisons with non-normal distributions can be found in the section on simulations. 
{\color{black}
We consider four local alternative hypotheses, along with the corresponding quantile effects  for the settings we present in the simulation section. 
\begin{enumerate}
    \item Location shift: $\beta(\tau) = \beta_n$, where $\beta_n = \beta_0/\sqrt{n}$.
    \item Location-scale shift: $\beta(\tau) = \beta_{1n}+\beta_{2n}F^{-1}(\tau)$, where $\beta_{1n} = \beta_1/\sqrt{n}$, $\beta_{2n} = \beta_2/\sqrt{n}$; $\beta_2 = \beta_1/2$.
    \item Lehmann shift (upper tail):  $\beta(\tau) = F^{-1}(1-(1-\tau)^{\beta_n})-F^{-1}(\tau)$, where $\beta_n = 1+\beta_0/\sqrt{n}$. This is equivalent to comparing two distributions $F(x)$ and $G(x)$, where $G(x) = F^{\beta_n}(x)$.
    \item Lehmann shift (lower tail):  $\beta(\tau) = F^{-1}(\tau^{\beta_n})-F^{-1}(\tau)$, where $\beta_n = 1+\beta_0/\sqrt{n}$. This is equivalent to comparing two distributions $F(x)$ and $G(x)$, where $G(x) = F^{1/\beta_n}(x)$.
\end{enumerate}
A visualization for Lehmann shift (lower/upper quantile effects) can be found in the  Supplementary Material.}
\begin{figure}[!ht]
    \centering
    \includegraphics[height = 38mm, width = 60mm]{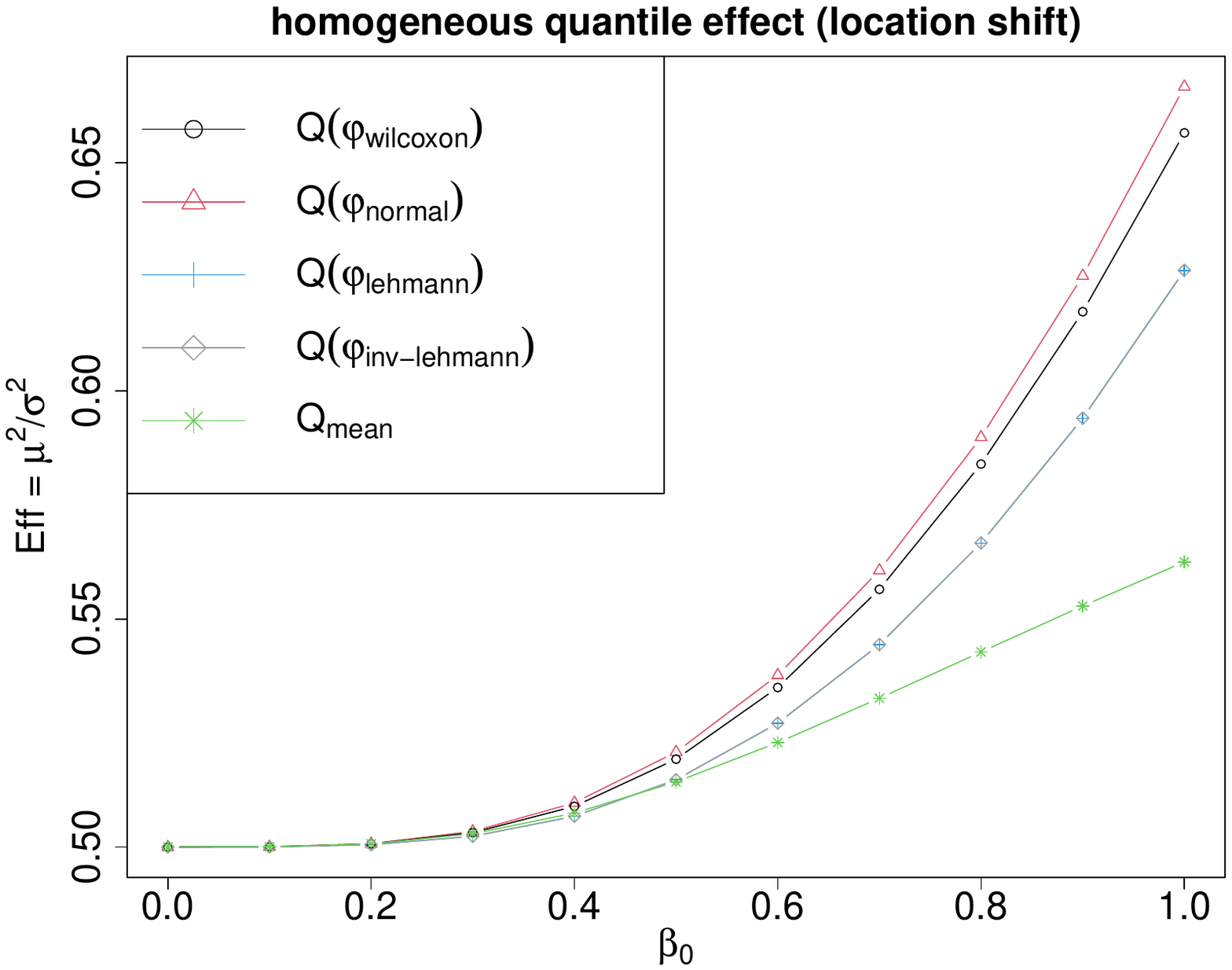}
       \includegraphics[height = 38mm, width = 60mm]{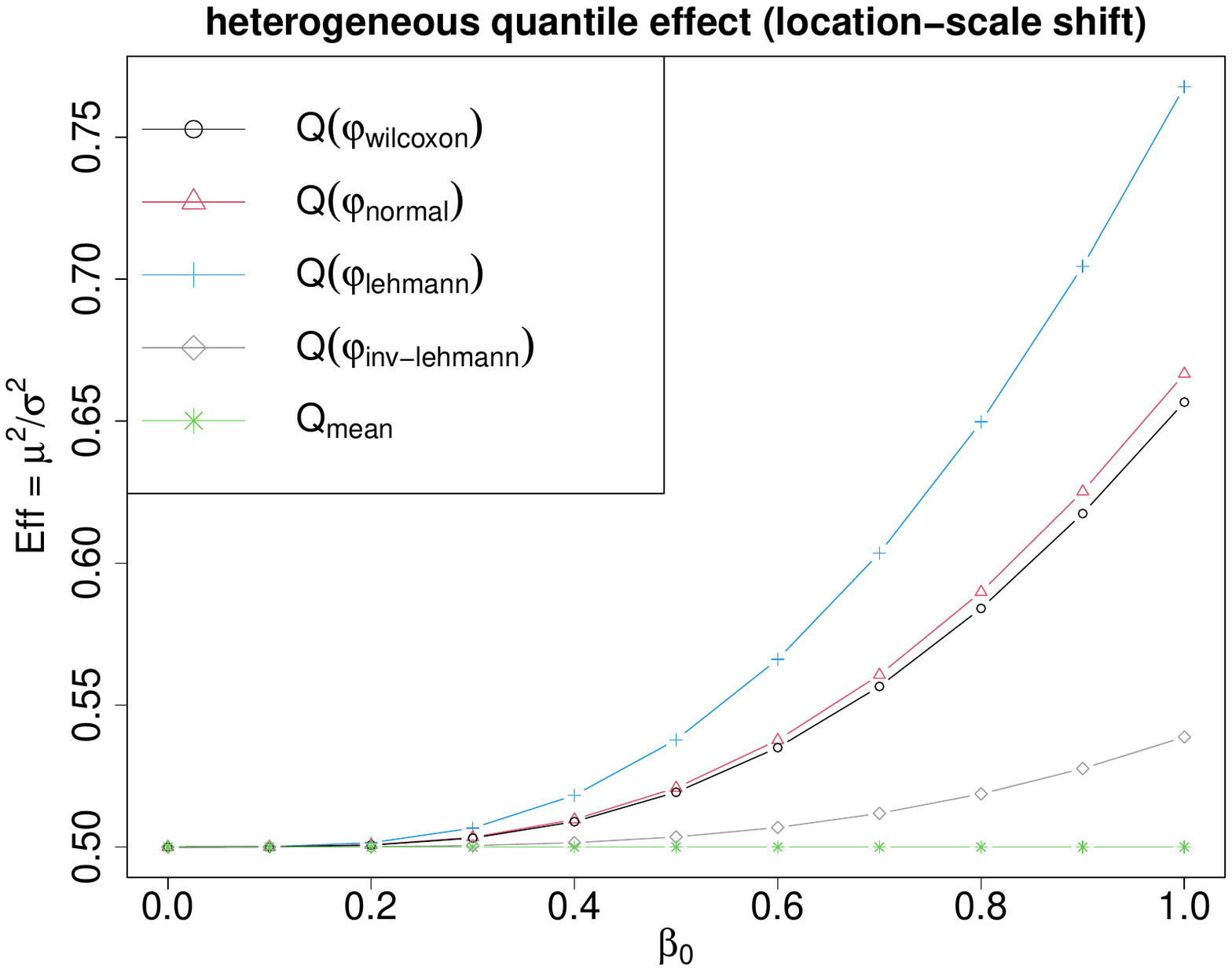}
       \includegraphics[height = 38mm, width = 60mm]{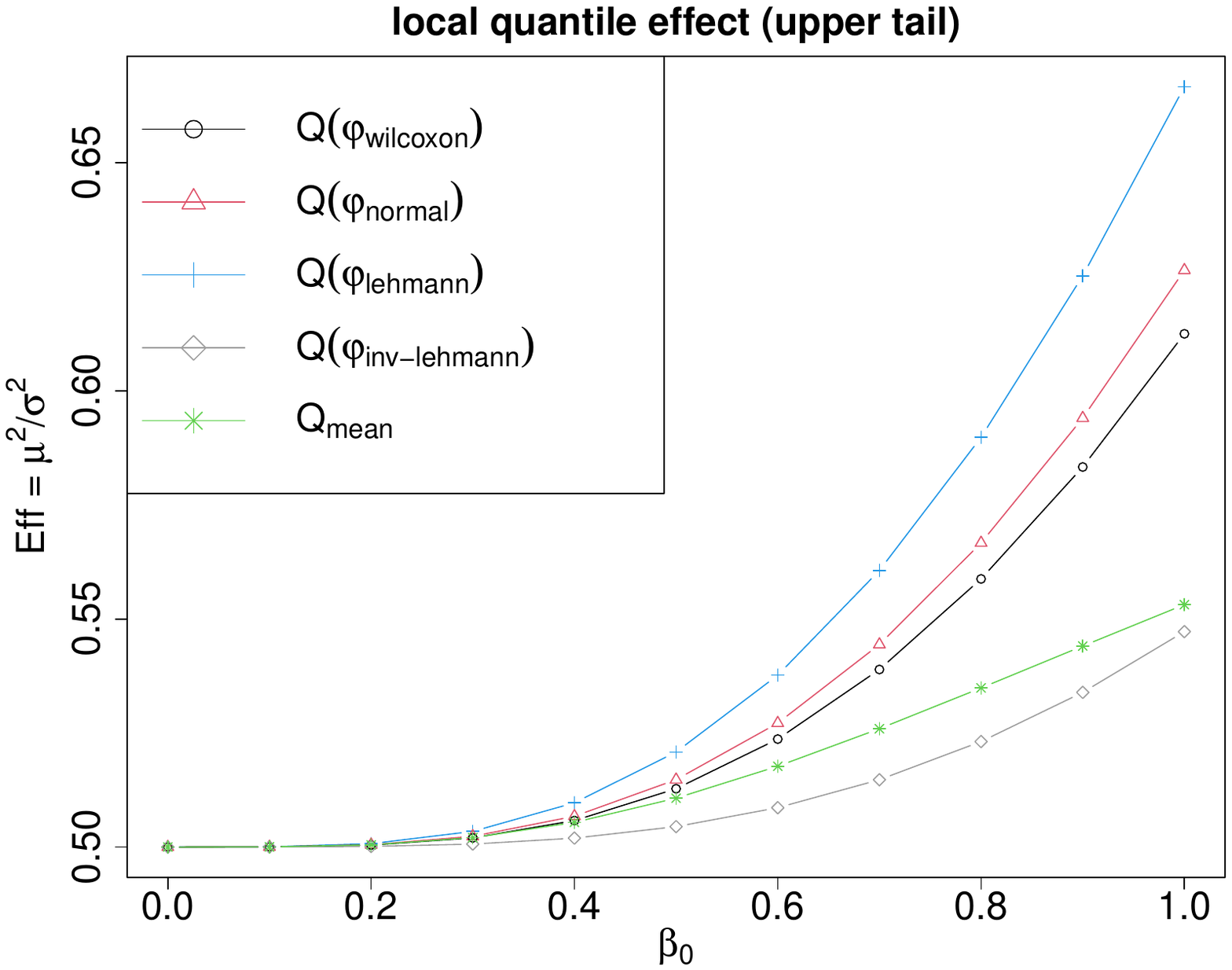}
       \includegraphics[height = 38mm, width = 60mm]{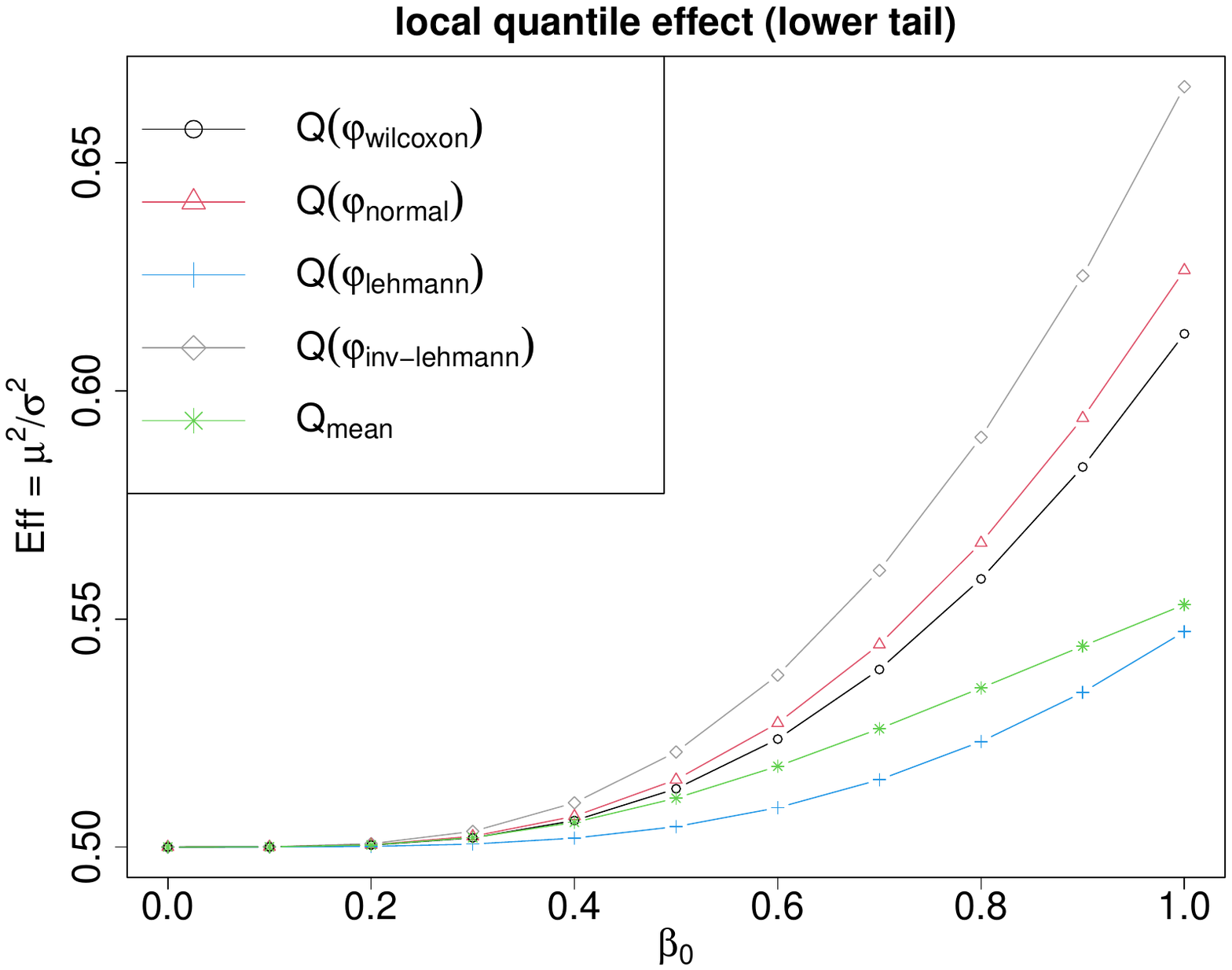}
    \caption{ Compare relative efficiency for different test statistics under the location shift (\textbf{top left}), location-scale shift (\textbf{top right}), upper-tail Lehmann alternatives (\textbf{bottom left}), and lower-tail  Lehmann alternatives (\textbf{bottom right}). In each figure, we present the relative efficiency for proposed test statistics $Q_\varphi$ based on four quantile weighting functions Wilcoxon, Normal, Lehmann and Inverse Lehmann, respectively.} 
    \label{fig:Eff}
\end{figure}
{\color{black}
Results are presented in Figure \ref{fig:Eff}. We observed that the Normal and Wilcoxon quantile weighting functions are more efficient than the others in the homogeneous case (location shift). The Lehmann quantile weighting function leads to higher efficiency in the location-scale shift model. For the Lehmann shift model, where the signals are concentrated in the upper tail or lower tail, the corresponding Lehmann function and Inverse Lehmann function are the best, as expected. %The performance we observed in location shift and Lehmann shift (upper tail) are consistent with \cite{gutenbrunner1993tests,koenker2010rank}. 
}

\section{Simulation}

\subsection{Simulation Models and Settings} \label{sec:simulsetting}
In this section, we present a simulation study to demonstrate the finite sample performance of the proposed tests under various genetic models and various trait distributions. We compare the proposed iQRAT statistics ($Q_S$ and $Q_B$) to the traditional mean-based tests. Specifically, we compare the proposed $Q_S$ test to the SKAT-C test proposed in \cite{ionita2013sequence}. SKAT-C is a test for the joint effects of rare and common variants.  We also compare the proposed $Q_B$ test to the Burden-C test in \cite{ionita2013sequence}. These SKAT and Burden tests were implemented via the function \texttt{SKAT\_CommonRare} in the \texttt{R} package \texttt{SKAT} \citep{SKATr}. {\color{black}In the following, we denote the unified $Q_S$ as iQRAT$_S$-mix, and denote $Q_S^\varphi$ with different quantile weighting functions $\varphi$ as iQRAT$_S$-W/-N/-L/-invL, corresponding to Wilcoxon/Normal/Lehmann/Inverse Lehmann function respectively. Similar notations are adopted for $Q_B$.}

 {\color{black} We simulate $(y_i,\X_i, C_i)$ with $i=1,...,n$ to reflect the complexities of real genetic association studies with sequencing data, including different genetic correlation structures, different directions of  effects, and assume sparsity of causal effects within a gene. In each simulated data, the genotype matrix $(\mathbf{X})$ is simulated from the \texttt{R} package SKAT \citep{SKATr} mimicking the data structure in sequencing studies. We use more than 100,000 Monte Carlo replicates.} The reference data set from \texttt{SKAT} package consists of 10,000 haplotypes over a 200kb region, including 3,845 variants. These haplotypes were simulated using a calibrated coalescent model (COSI, \citep{schaffner2005calibrating}), mimicking linkage disequilibrium structure in populations of European ancestry. For each simulated dataset, we randomly selected a roughly 3.5kb  region from the reference data, and treated it as the ``targeted" gene. We then generated individual genetic profiles $\X_i$ in that ``gene" by randomly drawing and combining two haplotypes from the  10,000 haplotypes.  Moreover, we calculate sample MAF of all the variants in each selected region/gene from the reference data, and randomly picked 20\% common variants and 30\% rare variants as causal variants. {\color{black} Additional simulations for different sparsity of effects are included in the Supplementary Material, as the results are similar to what have been presented in this section.}
{\color{black}According to the asymptotic theorem, under the alternative hypothesis, the power of iQRAT with single quantile weighting function depends on the alternative hypothesis, the error distribution and the  quantile weighting function itself. Hence, we consider different quantile models to generate the phenotype $Y_i$, and consider four error distributions, namely $N(0,1)$, $\chi^2_2$, $Cauchy(1,0)$, and $t_2$. }In all the models, we assume the covariate $C_i\approx N(4,1)$.

\begin{description}
\item[Global Model 1: a location model.] We assume that the phenotype $Y_i$ follows the model
$$Y_i = 1 + 1.2C_i+ \mathbf{X_i} \boldsymbol{\beta} + e_i \ \ i = 1,...,n, $$
where $\mathbf{X_i}$ is the vector of  genotypes for the $i$-th individual at the $k$ causal variant, $\boldsymbol{\beta}=(\beta_1,..,\beta_j,...,\beta_k)$,  $\beta_j = \beta |\log_{10}(m_j)|$ and $m_j$ represents the sample MAF of the $j$th causal variant. In this model, quantile effect is constant across all quantiles. We let $\beta = 0.3$ when the error distribution of $e_i$ is $N(0,1)$ or $\chi^2_2$, and let  $\beta = 0.6$ when the error distributions are $Cauchy(1,0)$ or $t_2$ with heavy tails and larger variation.

\item [Global Model 2: a location-scale model.] We simulated phenotype $Y_i$ from the following model,
\begin{equation}\label{model1}
Y_i = 1 + 1.2C_i+ \mathbf{X_i} \boldsymbol{\beta}+(1+\mathbf{X_i} 
\boldsymbol{\gamma})e_i,\ \ i = 1,...,n, 
\end{equation}
where $\beta_j = \beta |\log_{10}(m_j)|$, $\gamma_j = \gamma |\log_{10}(m_j)|$.  In this model, $\mathbf{X_i}$ is  associated with both the mean and the variance of $Y$.  We let $\gamma = 0.1$, $\beta = 0.3$ when the error distribution is $N(0,1)$ or $\chi^2_2$; and let $\gamma = 0.2$, $\beta = 0.6$ when the error distribution is $Cauchy(1,0)$ or $t_2$.

%\item [Model 3: a location-scale model with bi-directional associations.] Same model as in Model 2, but we randomly select 50\% of causal variants to have negative effects, while the others remain positive.

\item [Local Model: upper/lower quantile effect models.] In this setting, we assume that the conditional quantile function of the phenotype $Y$ can be written as 
$$Q_Y(\t | {\C}, {\X}) = 1 + 1.2\C + \X \beta(\tau)+ F^{-1}(\tau),$$ where $\beta(\tau) = 5\beta(\tau - 0.7)/(1-0.7)$ when $\tau>0.7$ and $\beta(\tau) = 0$ otherwise. Here the quantile effects only exist at the upper quantiles, i.e., $\tau \in [0.7,1]$. {\color{black}We also simulate a local model with lower quantile effects for $\tau \in [0,0.3]$ in a similar fashion.} Since the association only exists in a small interval, we set $\beta = 0.9\ /\ 1.8\ /\ 4.5\ /\ 2.7$ when the error distribution is $N(0,1)\ /\ \chi^2_2\ /\ Cauchy(1,0)\ /\ t_2$. To simulate $Y_i$ from this model, we use the inverse quantile approach, where we randomly draw a $U(0,1)$ random variable as $\tau$, and plug it into the conditional quantile function $Q_Y(\t | C_i, \X_i)$.
\end{description}

 After simulating $Y_i$ from these models, we use the quantile and rank normalization in \cite{qiu2013impact} to transform $Y_i$'s into a normal distribution.  Since quantile function is invariant to monotone transformations, the proposed iQRAT actually produces identical results with and without normalization. We implement the normalization for a fair comparison with the existing approaches. Otherwise, the existing methods will have type I error inflation issues, especially with non-gaussian errors.

% Ideally, we will independently simulate $N$ Monte-Carlo replicates of $\{\X, \textbf{C},\textbf{Y}\}$; $N$ depends on the significance level $\alpha$ we want to examine for the type I error. However, it is computationally heavy when $\alpha$ is small. To investigate whether iQRAT preserves the type I error rate at exome-wide threshold level $\alpha = 6.25e$-6, it requires conducting simulations on at least $10^7$ datasets. To reduce the computation Burden, we took the same strategy as in \cite{wu2011rare}. We randomly select 1,000 sets of regions with average length 3.5kb, as the genotype matrix ($\X$). Then, for each of the 1,000 sets, we simulated 10,000 sets of covariates $\textbf{C}$ and phenotype $\textbf{Y}$, such that we obtained $10^7$ individual genotype-phenotype datasets. Note that the same strategy is adopted in power study with total $10^5$ genotype-phenotype datasets: we simulated 100 sets of $\textbf{C}$ and $\textbf{Y}$ for each of the 1000 sets of $\X$.

{\color{black} Due to limited space, we present here only iQRAT$_S$-related results, and report the results of iQRAT$_B$ in the Supplementary Material.}

\subsection{Type I Error}
We first investigate whether the proposed iQRAT ($Q_S$ and $Q_B$) tests preserve the desired type I error rate at significance levels $\alpha =5e$-02, $1e$-02, $1e$-03, $1e$-04, $1e$-05, and at the exome-wide significance level $2.5e$-06. To do so, we simulated the data with sample size $n = 1,000$ under the null model, where $\beta = \gamma = 0$ and $e_i\sim N(0,1)$ in Model (\ref{model1}).  We present in Table \ref{T:type1error} the resulting type I error for iQRAT$_S$ from $10^7$ Monte-Carlo replicates. As shown in Table \ref{T:type1error}, iQRAT$_S$ test statistics have controlled type I errors at all significance levels. The slight inflation at the exome-wide significance level $2.5e$-06 is still within the 95\% confidence interval. Similar results were found in other scenarios where the error terms $e$ follow non-Gaussian distributions. {\color{black} The type I error is also controlled  for iQRAT$_B$, see detailed results in the Supplementary Material.}

\begin{table}[!h]
							\begin{center}
						
							\begin{tabular}{l|llllll}
							
									\hline

				&$\alpha = 0.05$  &  $\alpha = 0.01$  & $\alpha =1e$-03   &$\alpha =1e$-04    & $\alpha =1e$-05      &   $\alpha =2.5e$-06\\
					\hline
iQRAT$_S$-mix&    0.051& 9.77e-03& 9.34e-04 &8.78e-05& 9.10e-06& 2.7e-06\\
iQRAT$_S$-W&     0.048& 9.01e-03& 8.34e-04& 7.13e-05& 6.90e-06& 1.7e-06\\
iQRAT$_S$-N&      0.050& 9.83e-03& 9.44e-04& 9.25e-05& 1.03e-05& 2.2e-06\\
iQRAT$_S$-L&     0.050& 9.90e-03& 9.86e-04& 1.09e-04 &1.15e-05& 3.2e-06\\
iQRAT$_S$-invL&    0.050& 9.90e-03& 1.00e-03& 1.04e-04& 1.18e-05& 2.8e-06\\

%SKAT-C&   0.049& 9.57e-03& 9.37e-04& 9.86e-05& 9.80e-06 &2.3e-06\\
	\hline
				
									\end{tabular}
								\caption{\baselineskip=12pt Summary of Type I error for iQRAT test statistics. iQRAT-mix is the unified test statistic; iQRAT-W/-N/-L/-invL is the iQRAT using single quantile weighting function Wilcoxon/Normal/Lehmann/inverse Lehamnn. }\label{T:type1error}
							\end{center}
						\end{table}

\subsection{Power}
 We investigate and compare the empirical power of the proposed iQRAT test statistics and the competitors under the outlined model settings in Section \ref{sec:simulsetting}. We simulate data from each model setting with four different sample sizes $n = \{100, 500, 1000, 2000\}$.  We applied the proposed iQRAT tests, as well as the SKAT-C and Burden-C tests, to detect gene-level associations at the exome-wide significance level $\alpha = 2.5e$-06. We calculate the empirical power with $10^5$ Monte-Carlo replicates. {\color{black} We present the results of iQRAT$_S$ and SKAT-C with sample size $n=1000$ in this section, and present the results for other sample sizes in the Supplementary Material. Furthermore, the results for iQRAT$_B$ and Burden-C are also presented in the Supplementary Material.} 

\begin{figure}[!ht]
 \centering
\includegraphics[height = 38mm, width = 60mm]{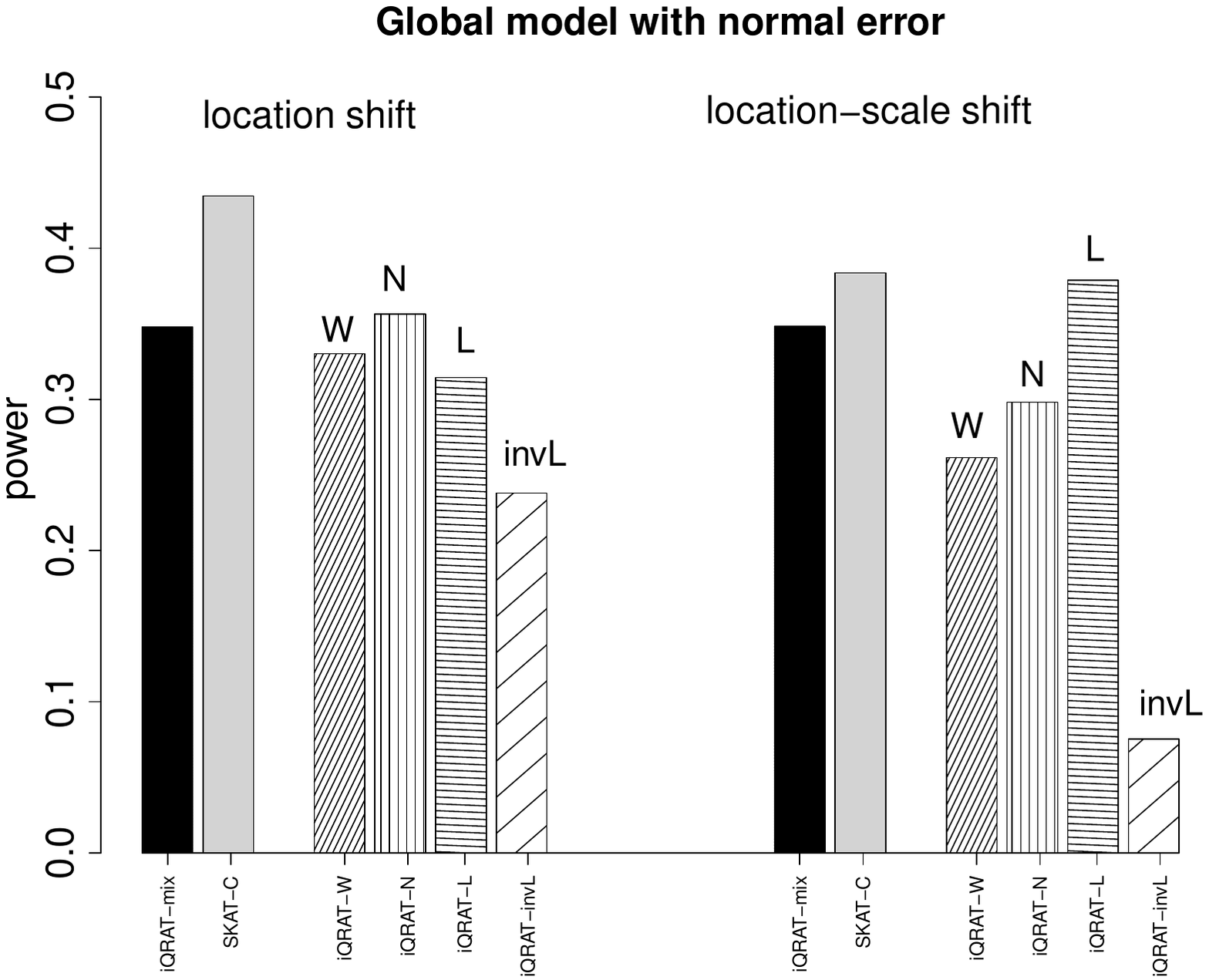}
\includegraphics[height = 38mm, width = 60mm]{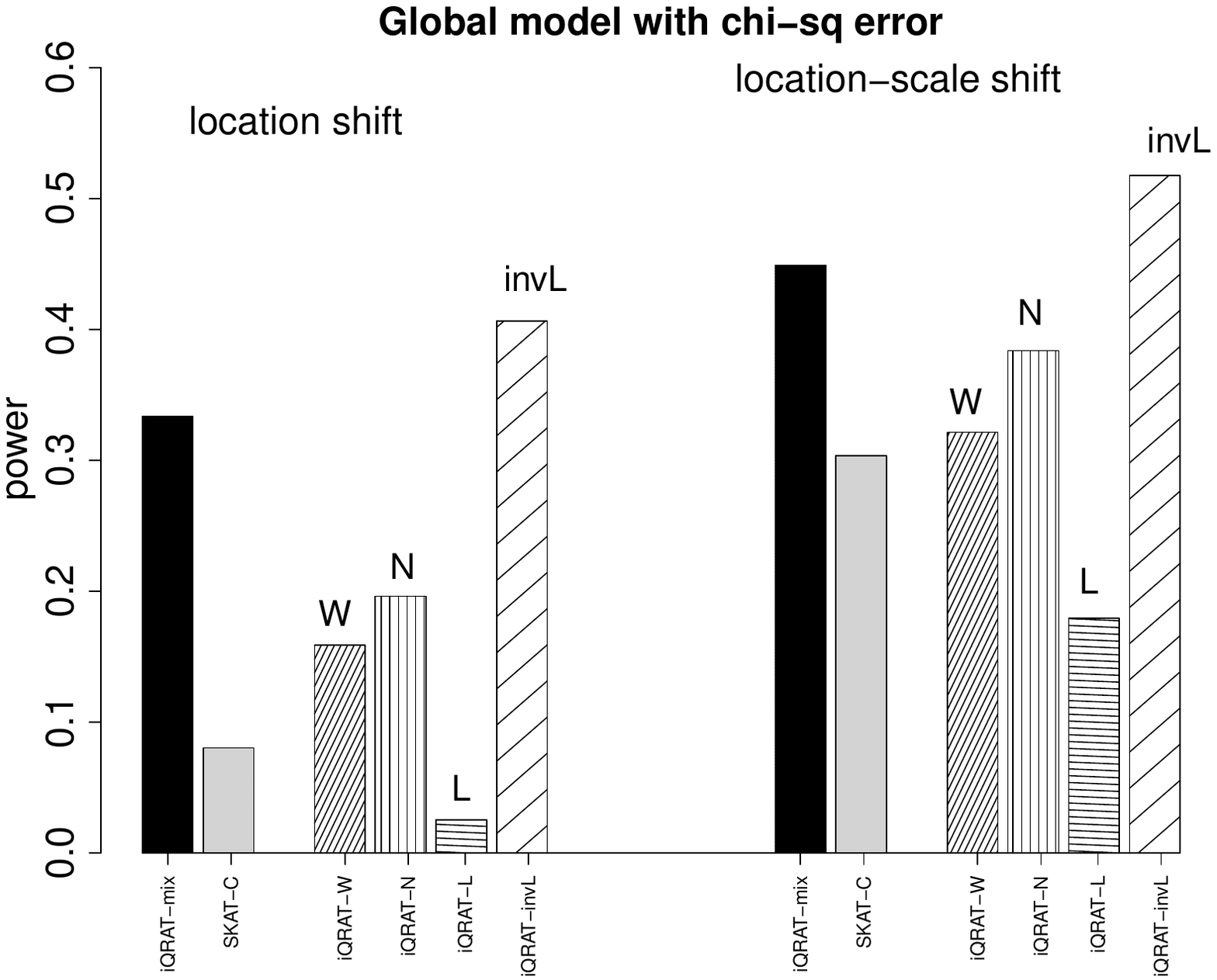}
\includegraphics[height = 38mm, width = 60mm]{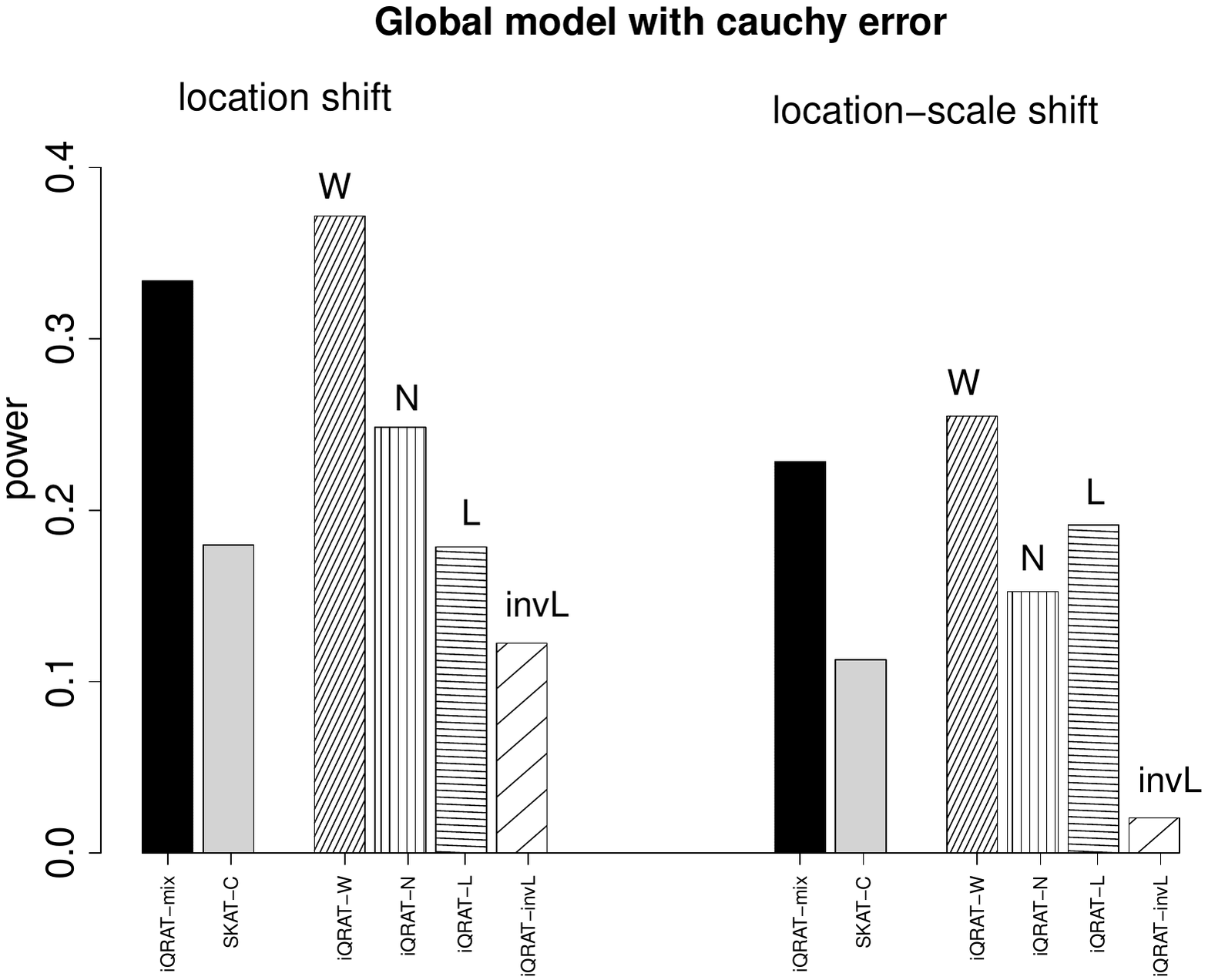}
\includegraphics[height = 38mm, width = 60mm]{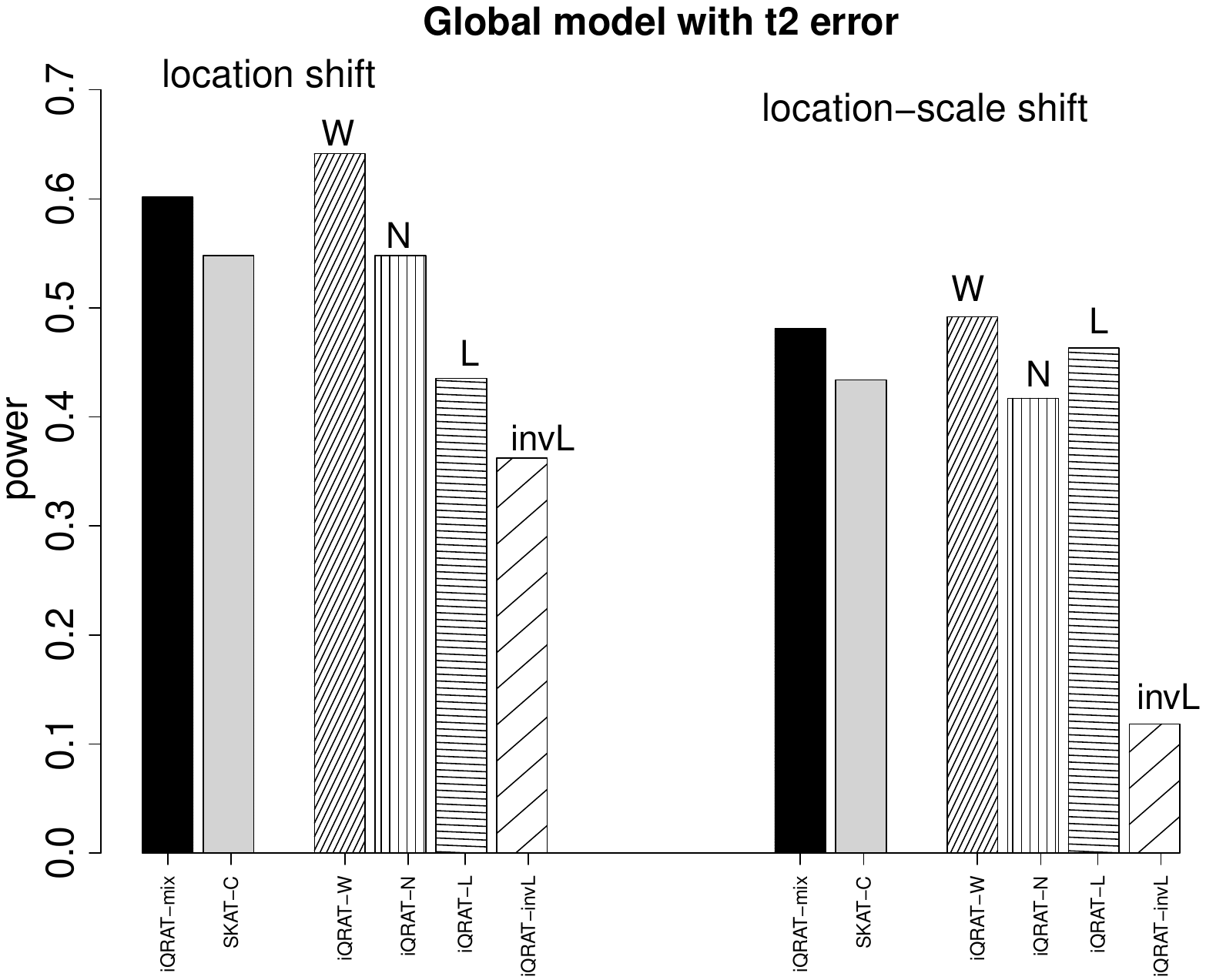}
\caption{Power results for iQRAT, and SKAT-C different scenarios, where causal variants are mix of common and rare variants. The significance level is 2.5e-06.}\label{fig:mix1000}
\end{figure}

\begin{figure}[!ht]
 \centering
\includegraphics[height = 38mm, width = 60mm]{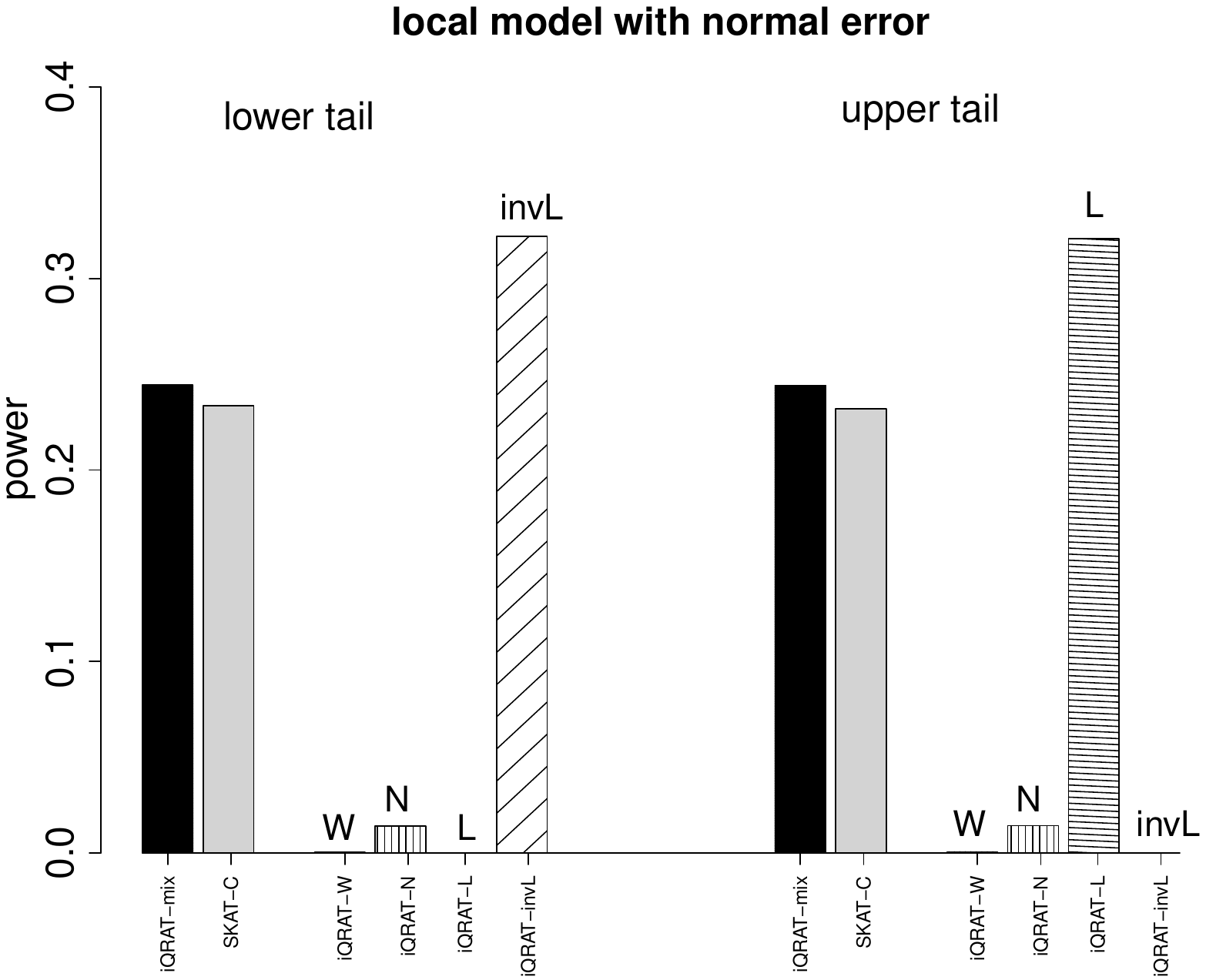}
\includegraphics[height = 38mm, width = 60mm]{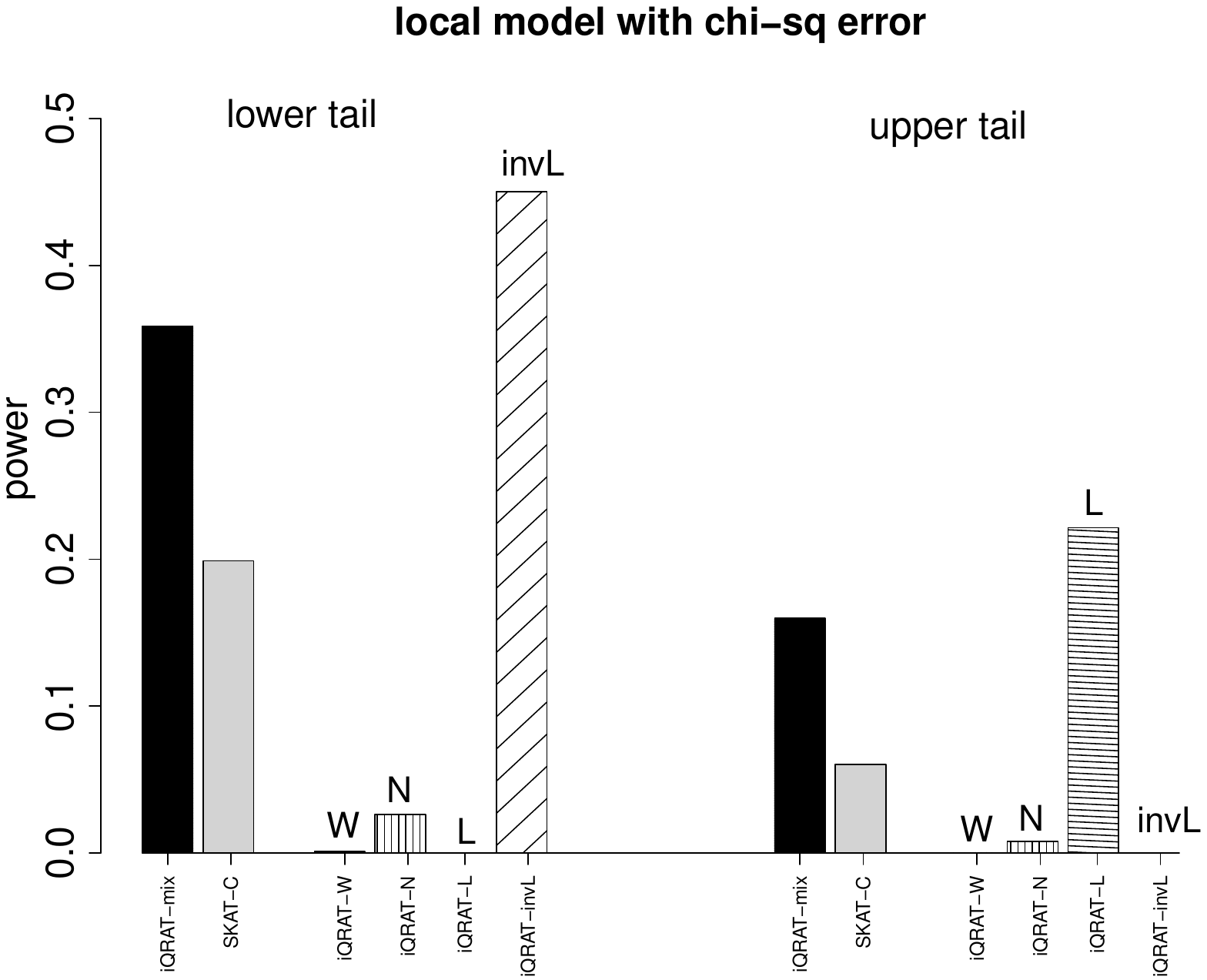}
\includegraphics[height = 38mm, width = 60mm]{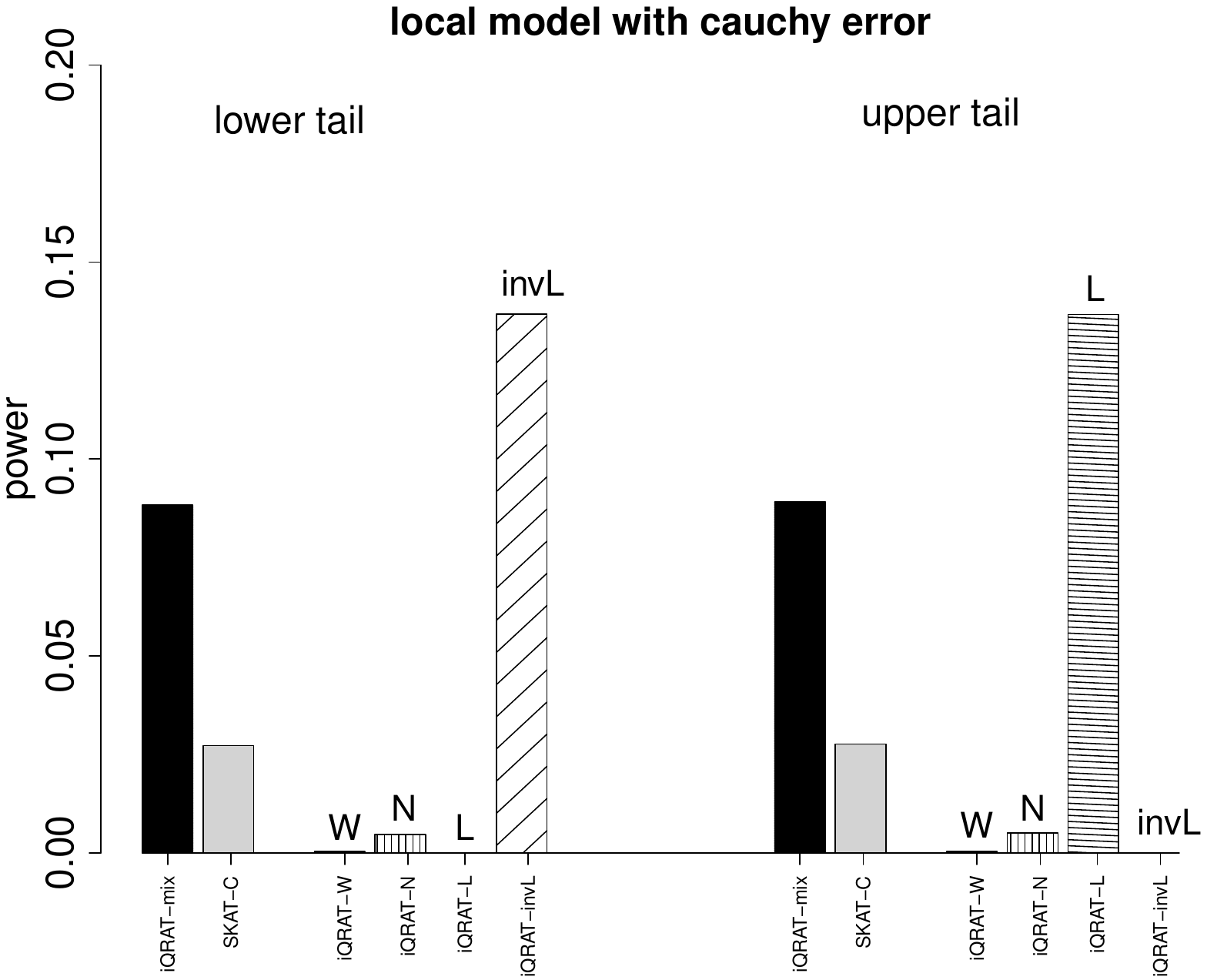}
\includegraphics[height = 38mm, width = 60mm]{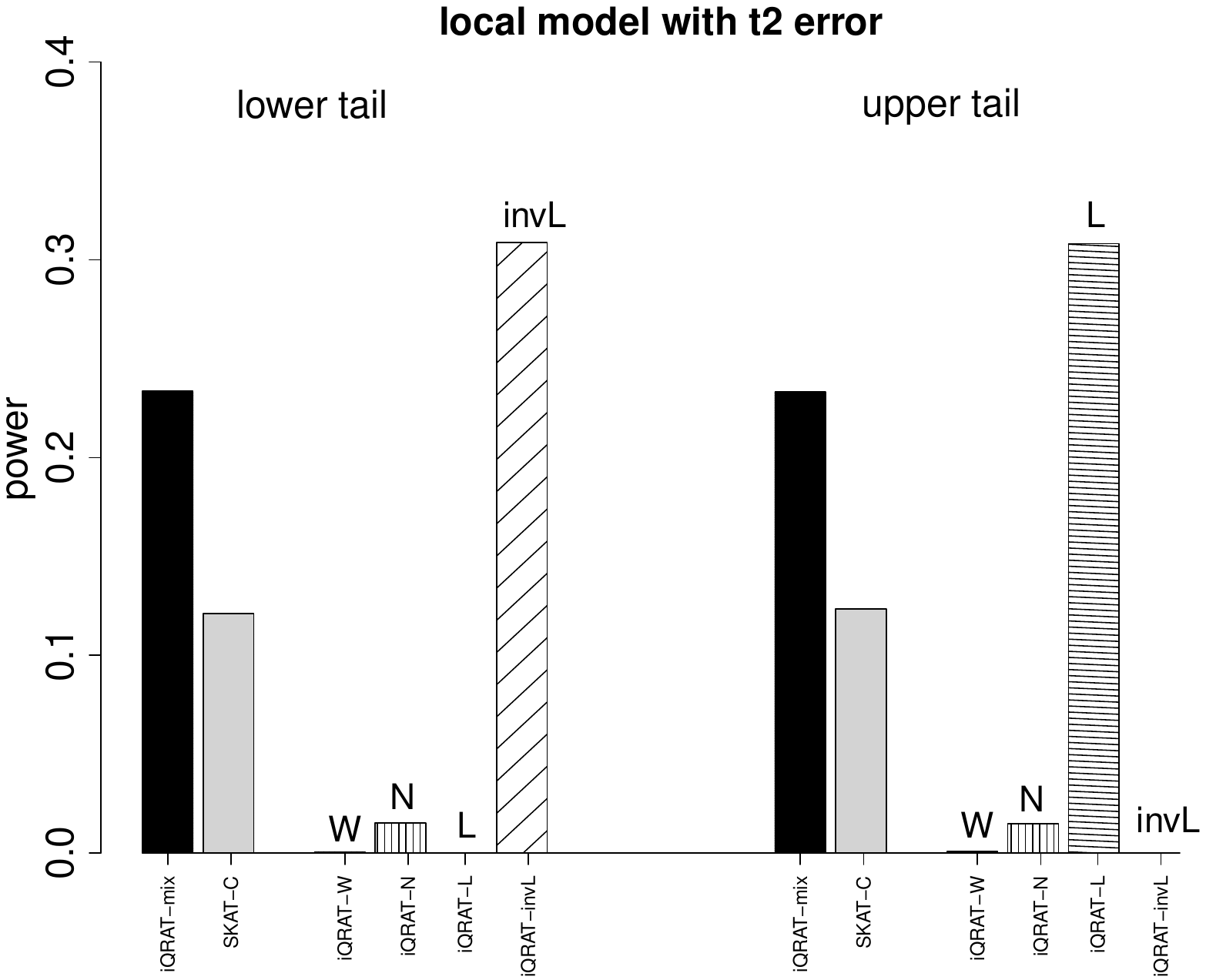}
\caption{Power results for iQRAT, and SKAT-C different scenarios, where causal variants are mix of common and rare variants. The significance level is 2.5e-06.}\label{fig:local}
\end{figure}

{\color{black}
We present in Figure \ref{fig:mix1000} the estimated power from $10^5$ Monte-Carlo replicates  under the  Global models 1-2 (i.e. Location shift and Location-scale shift). Each sub-figure corresponds to one specific error distribution. In each sub-figure and under each Global model, the first two bars represent the estimated powers from iQRAT$_S$-mix (the black bar) and SKAT-C (the light gray bar). The following four bars  represent iQRAT$_S$ with single quantile weighting functions (Wilcoxon/Normal/Lehmann/Inverse Lehmann), which provide insights for the power improvement of iQRAT$_S$-mix in different scenarios. Under the location model (i.e. homogeneous association) with normal errors, SKAT-C is slightly more powerful than the iQRAT as expected. When the error distribution is non-normal, iQRAT outperforms SKAT-C even after trait normalization. In the second Global model (i.e. the location-scale model), iQRAT and SKAT-C have comparable performance when the errors are normally distributed. When the errors are non-normal, the iQRAT again outperforms the SKAT-C. 
As we shown in the Supplementary Material, the efficiency gains are more evident as the sample size increases. 

The estimated power from the two Local Models are presented in Figure \ref{fig:local} with the same notations and legends.   When $X$ only impacts the tails of the $Y$ distribution, iQRAT outperforms SKAT-C under all error scenarios. As expected, the efficiency gain comes from the Lehmann/Inverse Lehmann weighted iQRATs, which upweight tail differences. Same as in the Global model,  the efficiency gains are more evident as the sample size increases (see the Supplementary Material). 

\subsection{Post-hoc analysis:}
Overall, we observed higher efficiency gain of the proposed iQRAT tests under heterogeneous genetic associations. Unlike in traditional quantile regression applications, here we do not have a target quantile level of interest. Thus, after identifying significant associations using the unified tests, we propose to perform post-hoc analyses via the single quantile weighting functions (iQRAT$_{\varphi}$ for specific weight function ${\varphi}$), to gain insights on gene-induced distributional differences, i.e. which part of the distribution has larger signals on gene-trait associations.  In this section, we found that such post-hoc analysis help better understand the power gain of iQRAT$_S$-mix in each scenario. 

We observed that iQRAT$_S$-W (Wilcoxon)  performs best for heavy-tailed errors such as the standard Cauchy and $t_2$ distributions; That is expected, and similar results were reported in \citep{zou2008composite}.
%\todo[inline]{cite hui zhou's work again} 
On the other hand, iQRAT$_S$-invL (Inverse Lehmann) had best power for $\chi^2$ errors. That is because the chi-square distribution has much higher density at the lower tail. Any location-shift effect will induce large detectable difference at lower tail.  When it comes to local models, iQRAT$_S$-L is clearly superior in detecting the local associations at the upper tail, while iQRAT$_{S}$-invL achieves the highest power for detecting the effect at the lower tail. By examining the patterns of p-values across the four different $\varphi$ functions, we also learn how the target gene affects the distribution of the phenotype $Y$. Such information   helps better identify the individuals at highest risk, and consequently leads to a more accurate risk stratification. 
}

% In all the analyses, we normalized the response $Y$ to follow the standard practice in genetic association tests. Such normalization only ensures the marginal distribution of $Y$ is approximately normal, but not the conditional distributions of $Y$ given $\X$ and $\C$. 

In reality, it is unlikely that all the conditional distributions of $Y$ given $\X$ and $\C$ are normally distributed. For this reason, we observed improved power with non-normal error distributions.

\section{Metabochip Data Analysis for Lipid Traits}
\subsection{Data Description}
The Metabochip is a custom genotyping array that assays nearly 200,000 variants in order to assess associations with traits such as type 2 diabetes, fasting glucose, coronary artery disease and myocardial infarction, low density lipoprotein cholesterol, high density lipoprotein cholesterol, triglycerides, body mass index, systolic and diastolic blood pressure, QT interval, and waist-to-hip ratio adjusted for BMI, etc \citep{voight2012metabochip}. In this section, we applied the proposed iQRAT on a Metabochip dataset  focusing on 265 genes in 99 gold fine-mapping regions. The data contain 12,281 individuals from eight studies, including  FUSION stage 2 ($n=2,741$), D2D 2007 ($n=2,108$), DPS ($n=429$), METSIM ($n=1,439$) and DR's EXTRA ($n=1,242$) in Finland;  HUNT and Troms\o ($n=2,793$ together) in Norway; and DIAGEN ($n=1,529$) in Germany. The two Norwegian cohorts are analyzed jointly as in \cite{he2018semi}. As a result, we have seven independent sites for the subsequent meta-analyses. We consider four lipid  phenotypes, low-density lipoprotein (LDL) cholesterol, high-density lipoprotein (HDL) cholesterol, total cholesterol (CHOL) and triglycerides (TG). {\color{black} We present the results for TG in the main manuscript, and report the results for HDL/LDL/CHOL in the Supplementary Material.} 
 We have excluded samples and SNPs with call rates $<98\%$, and also excluded any incomplete data with missing outcomes or covariates. The missing values in genotypes were imputed using mean imputation. Same as in the simulation studies, we compared the results to SKAT-C. In all the tests, we have adjusted for the covariates gender, age, squared age, and type 2 diabetes status for each study. For METSIM, we did not adjust for gender because it contains males only. For the two Norwegian studies, we additionally adjusted for study region. We did not adjust for principal components accounting for ancestry, because the Metabochip data is targeted array rather than genome-wide. The adjustments mentioned above are consistent with  \cite{lee2013general} and \cite{he2017unified}.

\subsection{Results}

Following the procedure in \cite{he2018semi}, we first apply the different tests to each study site, and then use Fisher's method \citep{fisher1992statistical} to combine the $p$-values across the seven sites.  {\color{black} As in the simulation study, we have compared iQRAT$_S$ to SKAT-C in the main text, and presented iQRAT$_B$ and Burden-C in the Supplementary Material. For each testing method, we use the exome-wide significance level $\alpha = 2.5$e-06. Both iQRAT$_S$-mix and SKAT-C identified six significant gene-trait associations exceeding the exome-wide significance level. The $p$-values from both tests as well as the single weighted iQRATs are listed in Table \ref{T:TG_S}.  Although iQRAT$_S$-mix and SKAT-C found the same number of exome-wide significant genes, the p-values from iQRAT$_S$-mix are much smaller than those from SKAT-C with one exception. Furthermore, examining the patterns of $p$-values, we found that for the genes ZPR1, LPL and BUD13, Lehmann-weighted iQRAT gave the smallest p-values, while the inverse-Lehmann weight reported the largest $p$-values. This suggests that the gene-trait associations are stronger at upper tails than at the lower tails.  And that is consistent with the empirical evidence for heterogeneous gene-trait associations shown in Section \ref{sec:intro}. 

Evidence in literature also supports that LPL is a well-known triglyceride-lowering gene, which  plays a critical role in breaking down fat in the form of triglycerides \citep{ference2019association}. For ZPR1, as effects are evident across quantiles, all tests reported significant $p$ values. This association has also been confirmed in the literature \citep{ueyama2015association,justice2018direct}.

}

\begin{table}[!h]
							\begin{center}
						
							\begin{tabular}{l|ll|llll}
							
									\hline

			Gene	&iQRAT$_S$-mix &   SKAT-C &  iQRAT$_S$-W   & iQRAT$_S$-N   &iQRAT$_S$-L  & iQRAT$_S$-invL    \\
					\hline

ZPR1  & 4.13e-47 & 3.26e-35&1.44e-41 &4.76e-46& 1.91e-50& 3.40e-28\\
LPL   & 1.70e-18& 2.97e-15& 4.70e-14 &2.66e-15& 8.85e-19& 1.05e-10\\
BUD13 & 1.59e-21& 1.38e-15& 3.42e-20 &1.51e-21 &4.50e-24 &2.83e-13\\
GCKR   &5.49e-08 & 3.13e-09&1.81e-07 &4.49e-08& 1.00e-08& 8.80e-08\\
ZNF512 &1.32e-06& 9.71e-06 &5.34e-06 &2.56e-05& 2.65e-07& 4.30e-07\\
MLXIPL &3.83e-07& 1.82e-06& 9.15e-08 &5.78e-07& 1.33e-06& 1.30e-06\\
						\hline
				
									\end{tabular}
								\caption{\baselineskip=12pt Meta-analysis results for gene-trait association test with respect to Triglycerides in Metabochip data. Exome-wide significant threshold 2.5e-06 has been applied.   }\label{T:TG_S}
							\end{center}
						\end{table}

 {\color{black}
 To obtain more insights into the heterogeneous gene-trait associations, we considered two ways for post-hoc visualization of the quantile-specific associations. 
 
 One way is to fit a semi-parametric quantile model $Q_Y(\tau|S_i,C_i) = C_i^\top\alpha_\tau + g_\tau(S_i)$, where $S_i$ is the aggregated mutation burden of the $i$th subject for the target gene (as defined in the motivating example in Section 1), and $g_\tau(S)$ is a non-parametric function of $S$. $g_\tau(\cdot)$ can be approximated by B-spline.  Based on the estimated model, we can estimate the quantile functions of $Y$ given
 the mutation burden $S$ at its 10th and 90th percentiles. Bootstrapping can be used to construct the corresponding confidence bands. Using the same Norwegian data and the two target genes (i.e., LPL and ZPR1) as in the motivating examples, we fit this semi-parametric quantile model, and show in Figure \ref{fig9} the estimated quantile functions with their 95\% bootstrap confidence band given the mutation burden of LPL  and ZPR1 at the 10th and 90th percentiles. 
 
 Higher mutation burden in LPL lowers the upper quantile of TG when $\tau>0.6$, while leaving the rest of the distribution unchanged.   For ZPR1, on the other hand, higher burden elevates the entire quantile function of TG, and the differences increase with quantile levels as well.  The horizontal doted line in Figure \ref{fig9} indicates the clinical suggested threshold for high levels of TG, i.e., 2.3mmol/L\footnote{from https://www.mayoclinic.org/}.  When one projects the intersection point (between a quantile function and high 
cholesterol threshold ) on to $X$ axis (as shown by the vertical dotted lines), we can easily obtain the probability/risk of high cholesterol.  As shown in Figure \ref{fig9},  carrying LPL mutations reduces  the risk of high cholesterol by at least 5\%, and could reduce the risk of higher cholesterol even more. Such findings indicate the potential clinical relevance of the heterogeneous association we discover.

A second way to investigate heterogeneity across quantiles is to plot the quantile-specific $p$-values. The proposed iQRAT tests integrate the rank-score process into a single test to enhance the detection power. Once a gene is identified, one can also  calculate quantile-specific $p$-values for mutation burden scores. The bottom panel of Figure \ref{fig9} plots the quantile-specific $p$-values of LPL and ZPR1, which is consistent with the quantile differences displayed in the upper panel.   %\todo[inline]{\bf maybe merge the two figures since they are related? AGREE}

%We can also validate the local quantile effects through traditional quantile regression by fitting quantile regression $Q_Y(\tau | S, C) = S^\top \beta(\tau)+C^\top \alpha(\tau)$ for a specific $\tau\in(0,1)$, where $S$ is the aforementioned Burden Score. By checking the $p$ value reported from \texttt{quantreg}, we can identify local signals for LPL and global signals for ZPR1, see Figure \ref{fig6}, which is consistent with Figure \ref{fig9} and Table \ref{T:norway2}. 

\begin{figure}[!ht]
 \centering
 \includegraphics[scale = 0.2]{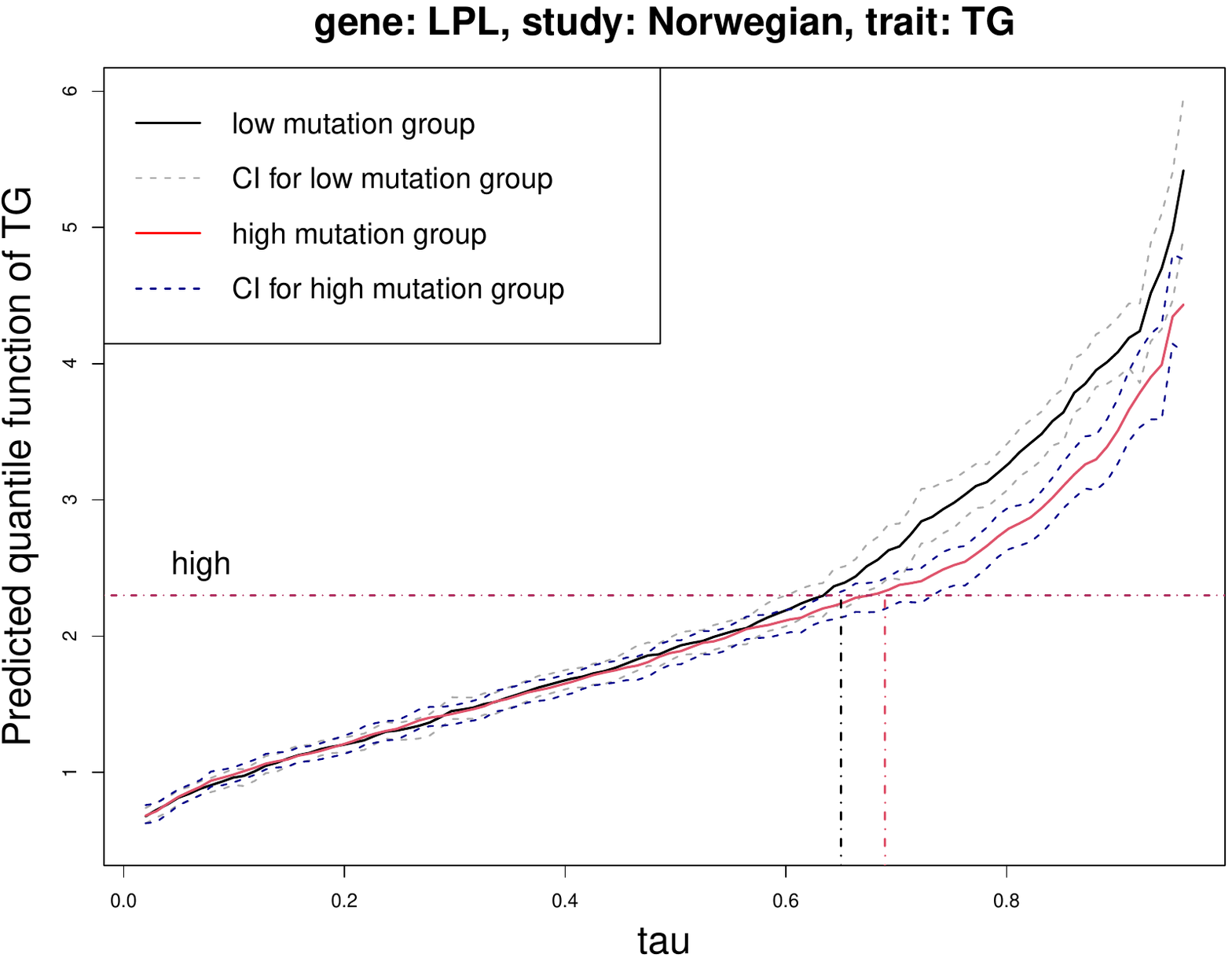}
 \includegraphics[scale = 0.2]{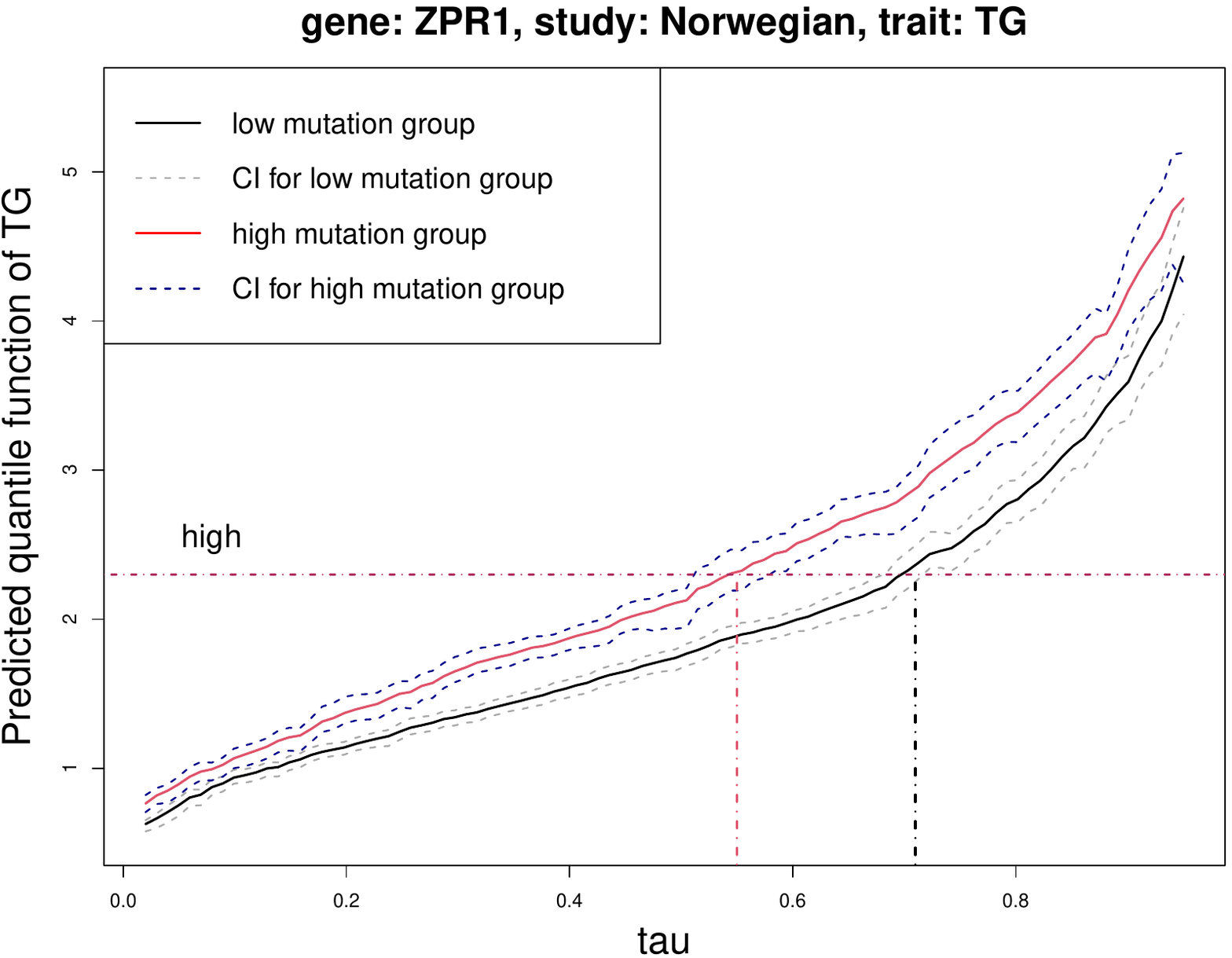}
%\caption{Predicted quantile function of $Y$ for  gene LPL (\textbf{left}) and gene ZPR1 (\textbf{right}). 95\% empirical confidence intervals are computed through bootstrap.}
 \includegraphics[scale = 0.2]{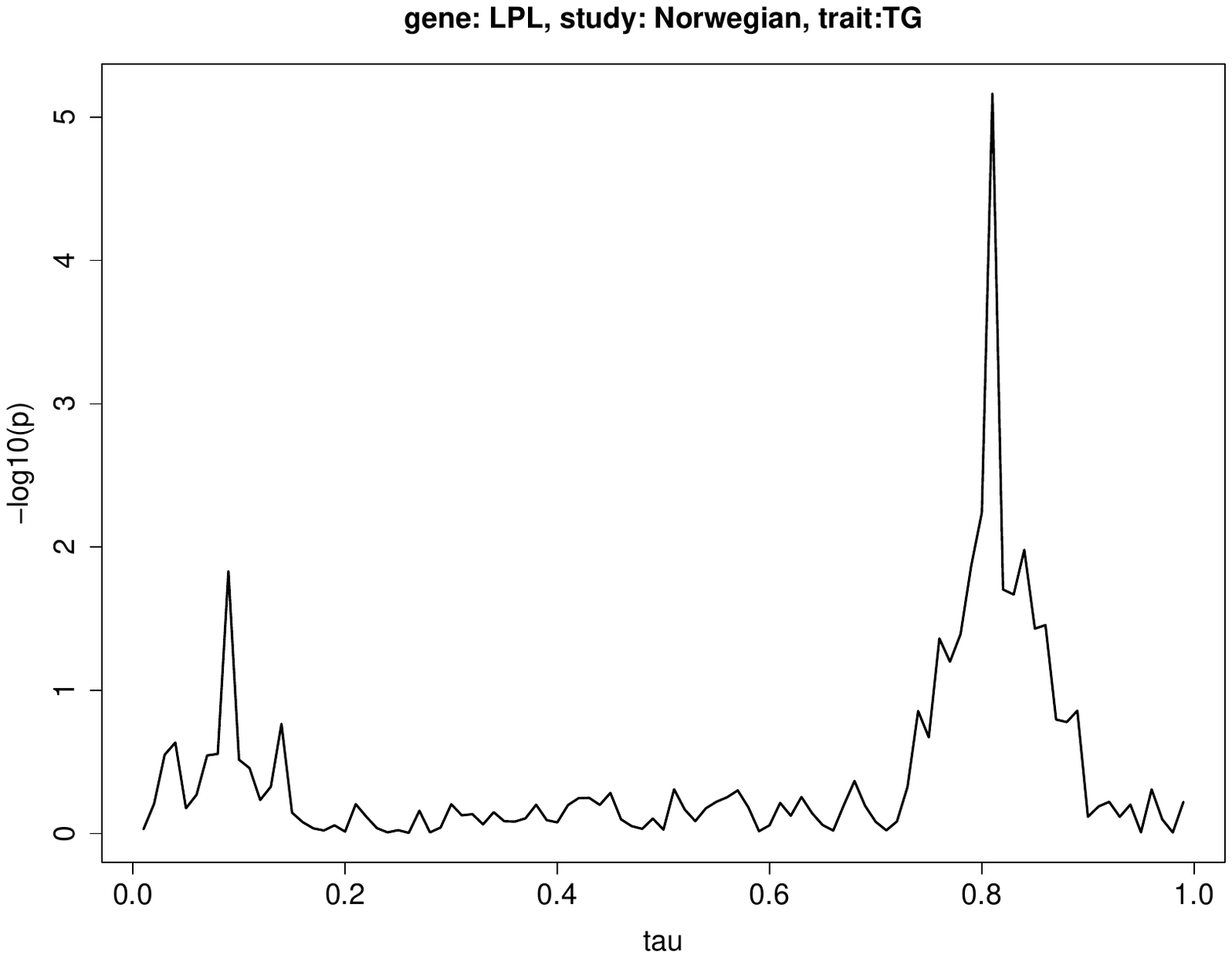}
\includegraphics[scale = 0.2]{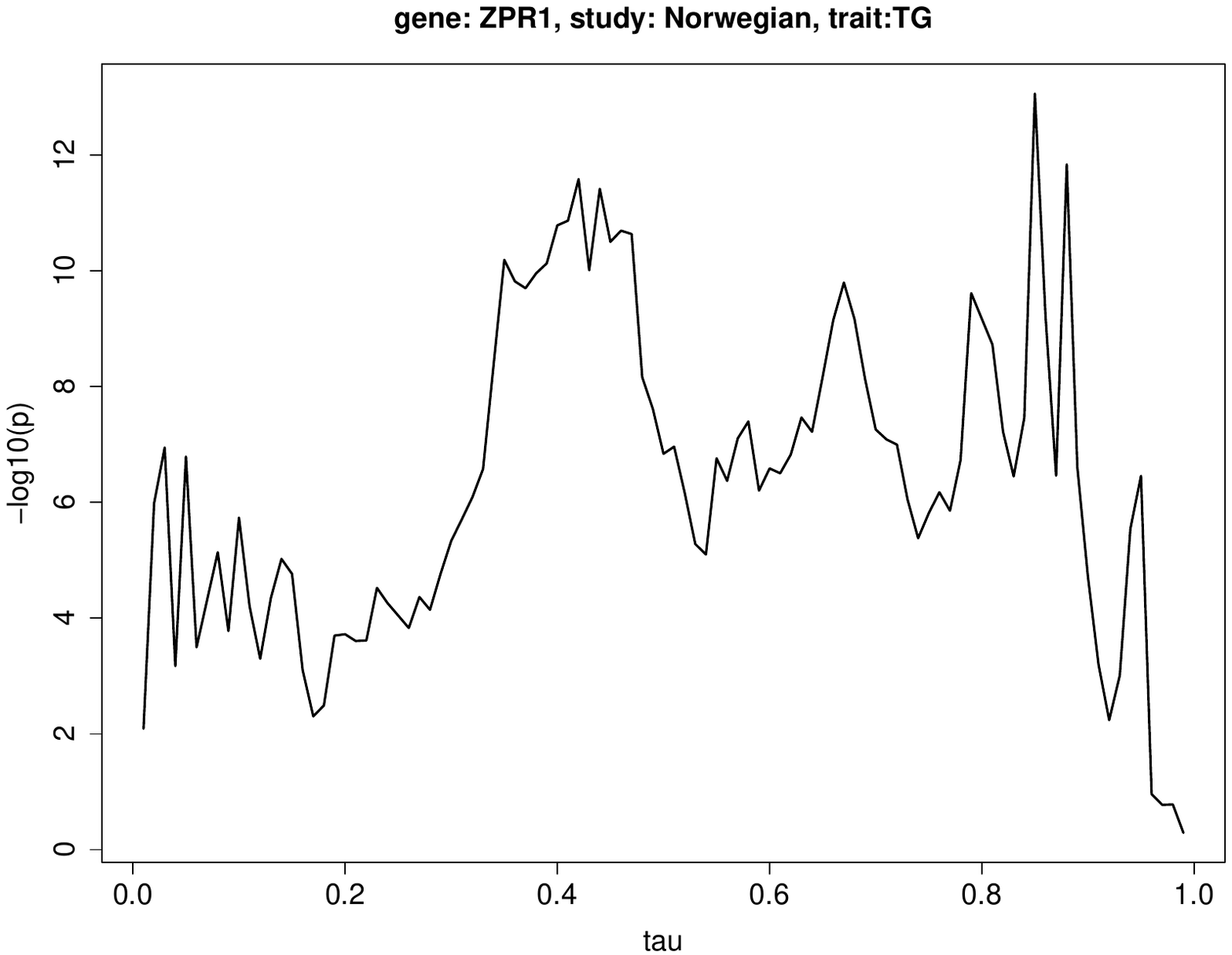}
\caption{\textbf{Top}: Predicted quantile function of $Y$ for  gene LPL (\textbf{left}) and gene ZPR1 (\textbf{right}). 95\% empirical confidence intervals are computed through bootstrap. \textbf{Bottom}: Validating local signals by traditional wald test for quantile regression. We report $\log_{10}(p)$ for gene LPL (\textbf{left}) and ZPR1 (\textbf{right}) in  Norwegian site, where $p$ is the p value of $\hat\beta(\tau)$ in $Q_Y(\tau | S, C) = S^\top \beta(\tau)+C^\top \alpha(\tau)$, for $\tau = \{0.01,0.02,...,0.98,0.99\}$.}\label{fig9}
\end{figure}
}

      \section{Discussion}
      
      In this paper, we propose an efficient integrated quantile test (iQRAT) based on weighted rank scores processes.  Compared to the widely-used mean-based dispersion and Burden tests, our test has the following advantages: (1) It is efficient and distribution-free.   By design, it is almost as efficient as the mean-based dispersion and Burden tests for homogeneous associations, and is more efficient in the presence of heterogeneous associations. Since the test statistic and the asymptotic distribution under the null are distribution-free, it is widely applicable to accommodates complex and heterogeneous associations. (2) Since quantile association is invariant to monotone transformation, it simplifies the data processing procedure by avoiding normalization, and enables direct interpretation on how a gene associates with the distribution/quantile functions of the phenotype.  Such insights are especially useful for exploring the genetic architecture of complex traits in more details. Moreover, avoiding trait normalization also facilitates  meta-analyses,  which is commonly performed in genetic analyses of multiple studies. Specifically, since the transformation functions used in normalization vary across individual studies, the summary statistics under different normalization procedures are not completely comparable from a technical perspective, which raises concerns when combining them across different studies. %I AM NOT SURE WE NEED TO INCLUDE THIS SENTENCE? {\color{black} Also, \cite{ray2020effect} investigated how non-normality of multivariate trait distribution affects the existing inference tools. They found that even with confounder controlled and residuals normalized for  each trait, some tests could still have inflated type I error for rare variants. Though we are not considering cross-phenotype association tests in this paper, it is desirable to make inference based on the original distribution of the trait, which in the future may shed light on developing association tests on the original multi-trait data directly.}      

        Although the proposed iQRAT test requires the estimation of the entire conditional quantile process, it is computationally feasible for large scale sequencing data due to the following reasons: (1) The estimation of quantile process uses the parametric linear programming technique that is much faster than estimating individual quantile functions. (2) The use of Cauchy combination to combine different weighting schemes is computationally simple. In the Metabochip data that we analyzed, iQRAT can be fully implemented within 1 second for testing a single gene in any of the eight studies, see the Supplementary Material for a summary of computational time.

        In the proposed iQRAT, we considered and combined four quantile weighting functions, each of them representing a different type of association. {\color{black}The Wilcoxon weight combines quantile effects equally across quantile levels, and is preferable for heavy tail error distributions. The Normal weight is heavier at the two tails and lighter in the middle range, and is optimal for normal errors. The Lehmann/Inverse Lehmann weight, on the contrary, assigns heavy weight at the upper/lower tails, and diminishes as quantile levels decrease/increase. They are designed to detect the right/left tail differences and location-scale changes.  By combining the various weighting schemes, the proposed unified iQRAT could support a wider range of complex and heterogeneous associations. After screening out possible associations, one can further consider using iQRAT with single quantile weighting function to detect heterogeneous associations.} Depending on individual applications, other weighing functions can be used without changing the asymptotic theory.

      Instead of pre-determined weight functions, it is also of interest to consider adaptive weights that may accommodate more complex associations. {\color{black}In \cite{ionita2013sequence},  an adaptive version of SKAT-C is proposed as SKAT-A, which combines the test statistics from common and rare variants in a more adaptive way. Considering data-driven combination of common and rare variants in iQRAT may lead to more informative discovery of complex gene-trait associations.} One could consider a two-stage procedure which estimates the quantile specific effects first, and then incorporates it into an integrated test. Implementation of such more adaptive integrated tests with appropriate type I error control warrants future research.

      In our application to the Metabochip data, most of the significant associations identified by iQRAT have already been identified using the classical SKAT-C tests.  This is expected since in general we only expect a small proportion of associations with higher order moments of the trait distribution, and we focused here on a small number of genes in 99 fine mapping regions. It is therefore of interest to apply the proposed methods to the genome-wide setting and multiple phenotypes to fully benefit from the power improvements we have shown in the simulations.

        \section*{R package}
      The proposed method has been fully implemented in the \texttt{R} package \texttt{iQRAT}, available  on Github: https://github.com/tianyingw/iQRAT. We will submit the package to CRAN once the paper is accepted. 
      
      \section*{Acknowledgments}
      We gratefully acknowledge Dr. Michael Boehnke for sharing the Metabochip data and  valuable insights. We also thank the investigators and participants from the FUSION, METSIM, HUNT, Tromso, DIAGEN, D2D-2007, DPS, DR's EXTRA studies that contributed data to the Metabochip data set. This work was supported by the National Institutes of Health grants R01 HG008980 and MH095797 and by National Science Foundation DMS-1953527. In addition, we would like to thank the Editor, Associate Editor and two anonymous referees for their valuable comments and suggestions.

% AOS,AOAS: If there are supplements please fill:
\begin{supplement}[id=suppA]
 \sname{Supplement A}
 \stitle{Integrated Quantile RAnk Test (iQRAT) for gene-level associations}
\slink[doi]{10.1214/00-AOASXXXXSUPP}
\sdatatype{.pdf}
 \sdescription{We provide additional material of (1) the technical details for Theorems 1 and 2;  (2) illustration of Lehmann alternatives and quantile effect of rank normalized trait; (3) comparison of computational times; (4) additional simulation results; (5) additional plots and tables for the meta-analysis of the Metabochip data.}
\end{supplement}

\bibliographystyle{imsart-nameyear}
\bibliography{01-bib-revision}

\end{document}